\documentclass[a4paper,11pt]{article}
\pdfoutput=1
\usepackage{jcappub}
\usepackage[utf8]{inputenc}
\usepackage[english]{babel}
\usepackage{physics}
\usepackage{amssymb,amsmath,amsfonts}
\usepackage{layout}
\usepackage[normalem]{ulem}
\usepackage{booktabs}

\usepackage{multirow}
\usepackage{verbatim}
\usepackage{comment}
\usepackage[usenames,dvipsnames]{xcolor}

\usepackage{graphicx,color}

\usepackage{orcidlink}

\usepackage{soul,float}

\usepackage{hyperref}

\newcommand{\cyc}{\textrm{cyc.}}

\makeatletter
\newcommand{\addbar@}[2]{%
  \makebox[0pt][l]{%
    \raisebox{#1}[0pt][0pt]{%
      \kern#2
      $\m@th\mathchar'26$%
    }%
  }%
}
\DeclareRobustCommand{\qbar}{\textrm{\addbar@{-1.5ex}{0.01em}}q}
\makeatother

\title{A novel, fast, and accurate code for cosmological loop calculations}

\author[a]{Tjark Harder,\orcidlink{0009-0006-9453-9841}}
\affiliation[a]{Institute for Theoretical Physics, Heidelberg University, Philosophenweg 16, 69120 Heidelberg, Germany}

\author[a,g]{Luca Amendola,\orcidlink{0000-0002-0835-233X}}

\author[b]{Thiago S. Pereira,\orcidlink{0000-0002-6479-364X}}
\affiliation[b]{Departamento de Física, Universidade Estadual de Londrina, Rod. Celso Garcia Cid, Km 380, 86057-970, Londrina, Paraná, Brazil}

\author[c, d, e, f]{Miguel Quartin\orcidlink{0000-0001-5853-6164}}

\affiliation[c]{Centro Brasileiro de Pesquisas Físicas, 22290-180, Rio de Janeiro, RJ, Brazil}

\affiliation[d]{Instituto de Física, Universidade Federal do Rio de Janeiro, 21941-972, Rio de Janeiro, RJ, Brazil}

\affiliation[e]{Observatório do Valongo, Universidade Federal do Rio de Janeiro, 20080-090, Rio de Janeiro, RJ, Brazil}

\affiliation[f]{PPGCosmo, Universidade Federal do Espírito Santo, 29075-910, Vitória, ES, Brazil}

\affiliation[g]{The Center for Astrophysics and Space Science (CASS), NYUAD Research Institute, New York University Abu Dhabi, PO Box 129188, Abu Dhabi, UAE} 

\emailAdd{harder@thphys.uni-heidelberg.de}
\emailAdd{l.amendola@thphys.uni-heidelberg.de}
\emailAdd{tspereira@uel.br}
\emailAdd{mquartin@cbpf.br}

\abstract{We introduce \texttt{neckar}, a public code for the one-loop matter and galaxy power spectrum, as well as correlation function, in the Eulerian Effective Field Theory of Large-Scale Structure (EFTofLSS). It is based on the FFTLog expansion, and is designed for fast and accurate parameter inference from large-scale structure data, making it a powerful tool for high-dimensional Bayesian analyses of current and upcoming galaxy surveys. \texttt{neckar} combines several numerical and analytical improvements, including the separation between pre-computable and cosmology-dependent contributions, a semi-analytic treatment of the Alcock–Paczyński effect for power spectrum multipoles, and an implementation of the symmetry-based bootstrap kernels, allowing straightforward applications to a broader class of cosmological models. The code also incorporates an improved infrared resummation scheme that corrects previous  
implementation inaccuracies, and which led to errors of the order of 1\% for the quadrupole, and of 10\% for the hexadecapole at scales \(k \gtrsim 0.1 \, h\textrm{Mpc}^{-1}\), as well as an asymptotic treatment that extends the range of accurate predictions into mildly non-linear scales \(k~\sim 0.4 \, h\textrm{Mpc}^{-1}\). This leads to relevant corrections in the correlation function around the BAO peak. Taken altogether, these improvements render \texttt{neckar} up to two orders of magnitude faster than similar existing codes without relying on emulators, and at the same time improving numerical accuracy. The code is written in C and is distributed with a Python wrapper for convenience.}

\begin{document}
\maketitle

\section{Introduction}\label{sec:intro}

The development of powerful and efficient Einstein-Boltzmann solvers, such as CAMB~\cite{Lewis:1999bs} and CLASS~\cite{Blas:2011rf}, together with high-precision measurements of the cosmic microwave background (CMB) from the Planck satellite~\cite{Planck:2018vyg}, has been the driving force behind some of the tightest constraints on cosmological parameters to date. In particular, it helped consolidate the $\Lambda$CDM model as the standard paradigm of cosmology almost a quarter of a century ago with the first precise all-sky measurements of the CMB with the WMAP probe~\cite{WMAP:2003elm}.

A complementary and potentially richer source of cosmological information lies in the large-scale structure (LSS) clustering of galaxies in the late-time Universe. While CMB and LSS surveys probe comparable ranges of physical scales, the CMB does so through a single two-dimensional projection on the last-scattering surface, whereas the LSS samples a three-dimensional volume; as a result, LSS surveys are expected to contain a few orders of magnitude more Fourier modes than the CMB, offering a correspondingly richer statistical harvest \cite{Ferraro:2022cmj,Ivanov:2022mrd,DESI:2022lza}.

Following decades of pioneering efforts to map the 3D distribution of galaxies through redshift surveys, from the early CfA Redshift Survey~\cite{deLapparent:1985umo} to the 2dF Galaxy Redshift Survey (2dFGRS)~\cite{2dFGRS:2005yhx} and the Sloan Digital Sky Survey (SDSS)~\cite{SDSS:2000hjo}, the modern Dark Energy Spectroscopic Instrument (DESI)~\cite{DESI:2016fyo} and the Euclid satellite~\cite{Euclid:2024yrr} are now able to produce very precise measurements. Probing the late-time Universe allows more stringent tests of the $\Lambda$CDM model, perhaps even leading to the discovery of new physics, as hinted by the combination of recent DESI results with supernova data, which suggest that dark energy may be evolving in time~\cite{DESI:2024mwx}.

Extracting cosmological information from clustering surveys traditionally requires efficient and accurate theoretical models of the clustering statistics: the $n$-point function in configuration, or polyspectra in Fourier space. Although recently the use of either field-level \cite{Shao:2022mzk,Andrews:2022nvv,deSanti:2023zzn} or simulation-based \cite{Tucci:2023bag,Zeghal:2024kic,FrancoAbellan:2024tbj} inference have drawn interest from the community, they incur specific challenges related to how accurate either the forward models or the simulations used in training are, and to the interpretability of the intermediate results.

As the leading two-point statistic, the power spectrum \(P(k)\) already captures a large fraction of the accessible information on sufficiently large, near-Gaussian scales. Building on Standard (Eulerian) Perturbation Theory (SPT)~\cite{Bernardeau:2001qr}, the Effective Field Theory of Large-Scale Structure (EFTofLSS)~\cite{Carrasco:2012cv} provides a renormalized perturbative description, and has manifested itself as one of the state-of-the-art tools for modeling the power spectrum in perturbation theory. Apart from galaxy bias and redshift space distortions (RSD), common ingredients also in SPT, the EFTofLSS readily extends the power spectrum already at one-loop order into the mildly non-linear regime via stochastic contributions, a systematic bias expansion and ultraviolet (UV) counterterms~\cite{Baumann:2010tm, Carrasco:2012cv, Senatore:2014vja, Senatore:2014eva, Senatore:2014via, Lewandowski:2015ziq, Perko:2016puo}.
At the same time, the FFTLog~\cite{Hamilton:1999uv} emerged as a tool for efficient computation of the EFTofLSS one-loop convolution integrals which compose the power spectrum, and facilitated Markov Chain Monte Carlo (MCMC) data analyses~\cite{McEwen:2016fjn, Simonovic:2017mhp}. In fact, it is already the standard for several codes, such as \texttt{FAST-PT}~\cite{McEwen:2016fjn, Fang:2016wcf}, $\texttt{CLASS-PT}$~\cite{Chudaykin:2020aoj}, and $\texttt{PyBird}$~\cite{DAmico:2020kxu} for Eulerian PT, as well as $\texttt{velocileptors}$~\cite{Chen:2020fxs, Chen:2020zjt} for Lagrangian PT. By replacing the numerically expensive one-loop integrals with fast semi-analytical evaluations of matrix multiplications, predictions of the power spectrum can be produced far more quickly without sacrificing accuracy.

These developments and tools have allowed, in recent years, the extraction of more information from  $P(k)$ using the one-loop corrections. Several  works have performed this type of analysis, many of them making use of the full-shape method and assuming a cosmological model, using data from different surveys: BOSS~\cite{Ivanov:2019pdj,Philcox:2021kcw,DAmico:2019fhj}, eBOSS~\cite{eBOSS:2020hur,Ivanov:2021zmi,Chudaykin:2022nru} and DESI~\cite{DESI:2024hhd}. Model-independent approaches have also been proposed which can still extract precise constraints from $P(k)$, such as the FreePower~\cite{Schirra:2024rjq} and ShapeFit methods~\cite{NovellMasot:2025fju}. In addition to the power spectrum, some analyses \cite{Philcox:2021kcw,Ivanov:2023qzb,Chudaykin:2025aux} have obtained tighter constraints through the inclusion of the tree-level bispectrum, which at this order requires no new perturbative ingredients beyond those already needed for the one-loop power spectrum.

In this work, a new implementation of the power spectrum and correlation function at one-loop order in the Eulerian framework of the EFTofLSS with the FFTLog method is presented that aims at improving other available codes in several ways. The code, \texttt{neckar}
(Non-linear Evolution with Complex Kernels for Accelerated Runs), is written in the C programming language, and relies on FFTW~\cite{Frigo:2005zln} and GSL~\cite{Galassi_2009} as external libraries. We also provide a Python wrapper for ease of use. The complete code will be made publicly available soon.\footnote{\href{https://github.com/tjharder/neckar}{https://github.com/tjharder/neckar}}

While existing codes based on the FFTLog already outperform conventional codes that rely on numerical integration techniques by several orders of magnitude in terms of computational time, a typical MCMC analysis is still computationally challenging. This is especially true for high-dimensional analyses involving many redshift bins, nuisance parameters, and repeated calls to Boltzmann solvers. In particular, apart from the few core parameters of a cosmological model, several nuisance parameters need to be explored in an MCMC run for each redshift bin in any given survey in order to account for the UV corrections, galaxy biases and stochastic terms. This problem is even more pressing in model-independent approaches like FreePower. Therefore, further reductions in runtime remain highly valuable.

This shortcoming is alleviated in \texttt{neckar} for several reasons. For one, the usual benefits of a compiled programming language over an interpreted programming language, such as Python, in which many of the available codes are written, apply. Furthermore, even more important, it is implemented in a very efficient way. 
But first and foremost, computations of contributions which are independent of the free parameters are separated from the rest. While the latter need to be updated at each step in the MCMC, the former, primarily consisting of numerically expensive tasks, only need to be computed once in the beginning. Finally, the inclusion of the Alcock-Paczyński (AP)~\cite{Alcock:1979mp} effect for the power spectrum multipoles is done in a novel, semi-analytic way, as will be explained in detail later on. Other codes, so far, relied on numerical integration techniques for the multipole decomposition of the AP corrected power spectrum.

As will be shown in detail below, excluding the overhead of generating initial conditions with Boltzmann solvers, these improvements generally lead to speed-ups of roughly two orders of magnitude compared to other codes, without relying on emulators or sacrificing numerical accuracy. This is enough to make CAMB or CLASS the potential bottlenecks for \texttt{neckar} by about one order of magnitude if  any of them need to be called at each step.

Lastly, two additional major novelties are included in the code. Instead of assuming the usual Einstein-de Sitter (EdS) approximation for the matter and velocity kernels, we rely on the symmetry-based kernels in the bootstrap approach of~\cite{DAmico:2021rdb}. This makes \texttt{neckar} more versatile, as several cosmological models can be explored by determining just a small number of parameters in the respective cosmology. Finally, the FFTLog compatible IR resummation introduced in \texttt{PyBird}~\cite{DAmico:2020kxu} is adopted while improving upon two main aspects: first, some inaccuracies have been corrected, which were resulting in relative errors on the order of $1 \%$ for the quadrupole and $10 \%$ for the hexadecapole on scales $k \gtrsim 0.1\, h\textrm{Mpc}^{-1}$. We will discuss the origin of these corrections below. Second, an asymptotic treatment is developed that extends the IR resummation accuracy deeper into the non-linear regime, while improving poor convergence of the IR resummation already at non-linear scales around $k \sim 0.4\,h\textrm{Mpc}^{-1}$ at the present epoch. As we will show, this strongly benefits the correlation function on scales around the baryon acoustic oscillations (BAO) peak, as ringing defects from the previous ad-hoc truncation are eliminated.

\section{Constructing the Power Spectrum}\label{sec:pk}

The power spectrum, $P(\vec{k}; z)$, being the Fourier transformed 2-point correlation function of the matter distribution in the Universe, is defined via~\cite{Peebles:1980yev}
\begin{equation}
    (2\pi)^3 \delta^{(3)}_D\big(\vec{k} + \vec{k}'\big) P\big(\vec{k}; z\big) = \left\langle \delta\big(\vec{k}; z\big) \delta\big(\vec{k}'; z\big) \right\rangle \, ,
\end{equation}
where $\delta^{(3)}_D(\cdot)$ is the 3-dimensional Dirac delta distribution, $z$ is the redshift, $\langle \cdot \rangle$ is the ensemble average, and $\delta(\vec{k}; z)$ is the Fourier transformed matter density contrast.

In presence of redshift-space distortions (RSD) in the plane-parallel approximation, apart from the redshift, the power spectrum is a function of $k = |\vec{k}|$ and $\mu = \hat{k} \cdot \hat{s}$ only, where $\hat{k} = \vec{k} / k$ and $\hat{s}$ is the line-of-sight (LOS)~\cite{Hamilton:1997zq},
\begin{equation}
    P\big(\vec{k}; z\big) = P(k, \mu; z)\,.
\end{equation}

Instead of using the full $\mu$-dependence of the power spectrum, it is common practice to use the multipoles of the power spectrum~\cite{Hamilton:1992zz},
\begin{equation}
    P^{\ell}(k) = \frac{2 \ell + 1}{2} \int_{-1}^{1} d\mu \, P(k, \mu; z) \mathcal{L}_{\ell}(\mu)\, ,
    \label{eq:Multipole}
\end{equation}
where $\mathcal{L}_{\ell}$ is the Legendre polynomial of order $\ell$, such that $P(k, \mu) = \sum_{\ell} P^{\ell}(k) \mathcal{L}_{\ell}(\mu)$.

Note that, in the following, we often omit the redshift dependency in the text when not directly necessary, i.e., we write $P(\vec{k}) \equiv P(\vec{k}; z)$.

\subsection{Alcock-Paczyński Effect} \label{sec:ap}

Mapping the 3D distribution of tracers, such as galaxies, from redshift surveys requires a reference cosmology to translate angles and redshifts $z$, into either comoving ($D_C$) or angular diameter ($D_A$) distances. If the true cosmology differs from this reference one, however, this will lead to distortions along and across the line of sight (LOS)~\cite{Ballinger:1996cd}. This effect is  known in the literature as the Alcock-Paczyński (AP) effect \cite{Alcock:1979mp} and needs to be accounted for in theoretical predictions of the correlation functions in configuration and Fourier spaces. 

With the notation of~\cite{Amendola:2022vte}, the power spectrum, $P(k, \mu)$, transforms under the AP effect as
\begin{equation}
    P'(k', \mu'; z) = \frac{h(z)}{d^2(z)} P\bigg(\alpha(\mu'; z) k', \frac{h(z)}{\alpha(\mu'; z)} \mu'; z\bigg) \, ,
    \label{eq:APEffectPowerSpectrumMu}
\end{equation}
where primed quantities are measured in the reference cosmology. We also define $h(z) \equiv H(z) / H'(z)$ and $d(z) \equiv D_A(z) / D_A'(z)$, the ratios of the Hubble expansion rate and the angular diameter distance, respectively, and
\begin{equation}
    \alpha^2(\mu; z) = \frac{1}{d^2(z)} \Big(1 - \eta(z) \mu^2\Big)
\end{equation}
with $\eta(z) \equiv 1 - [h(z) d(z)]^2$. While eq.~\eqref{eq:APEffectPowerSpectrumMu} is trivial when working with the $\mu$-dependent power spectrum, it becomes non-trivial if one wants to work with the power spectrum multipoles (cf. eq.~\eqref{eq:Multipole}),
\begin{equation}
    P^{\ell'}(k'; z) = \frac{2\ell' + 1}{2} \frac{h(z)}{d^2(z)} \int_{-1}^{1} d\mu' P\bigg(\alpha(\mu'; z) k', \frac{h(z)}{\alpha(\mu'; z)} \mu'; z\bigg) \mathcal{L}_{\ell'}(\mu') \, ,
    \label{eq:APEffectPowerSpectrumMultipole}
\end{equation}
where the notation $P^{\ell'}(k') \equiv (P')^{\ell'}(k')$ is used to indicate the power spectrum multipoles in the reference cosmology. Due to the non-trivial dependence of the power spectrum on \(k=\alpha(\mu';z)k'\), this integral is generally not analytically solvable, and instead requires the use of an expensive numerical integration technique. 

However, a more refined approach can be taken if the power spectrum is of the form 
\begin{equation}
    P(k, \mu; z) = \sum_m \mathcal{P}_m(\mu; z) \, k^{\tau_m} \, ,
    \label{eq:APEffectPowerSpectrumMuPowerLaw}
\end{equation}
or, when using the multipoles of the power spectrum, of the form
\begin{equation}
        P^{\ell}(k; z) = \sum_m \mathcal{P}^{\ell}_m(z) \, k^{\tau_m} \, ,
    \label{eq:APEffectPowerSpectrumMultipolePowerLaw}
\end{equation}
i.e., when the scale-dependence of the power spectrum can be separated from the angular dependency in the form of a (potentially complex) power series with exponents $\tau_m \in \mathbb{C}$. Note that the coefficients $\mathcal{P}_m(\mu; z) \in \mathbb{C}$ and $\mathcal{P}^{\ell}_m(z) \in \mathbb{C}$ are related via (cf. eq.~\eqref{eq:Multipole})\footnote{In the plane-parallel approximation, the power spectrum in redshift-space can be decomposed into a series in powers of $\mu^2$. The integral in eq.~\eqref{eq:APEffectPowerSpectrumCoefMultipole} can therefore be performed analytically.}
\begin{subequations}
    \begin{equation}
        \mathcal{P}_m(\mu; z) = \sum_{\ell} \mathcal{P}^{\ell}_m(z) \mathcal{L}_{\ell}(\mu) , \,
        \label{eq:APEffectPowerSpectrumCoefMu}
    \end{equation}
    \begin{equation}
        \mathcal{P}^{\ell}_m(z) = \frac{2\ell + 1}{2} \int_{-1}^1 d\mu \; \mathcal{P}_m(\mu; z) \mathcal{L}_{\ell}(\mu) \, .
        \label{eq:APEffectPowerSpectrumCoefMultipole}
    \end{equation}
\end{subequations}

Inserting eq.~\eqref{eq:APEffectPowerSpectrumMuPowerLaw} into eq.~\eqref{eq:APEffectPowerSpectrumMultipole} and using eq.~\eqref{eq:APEffectPowerSpectrumCoefMu}, the transformation under the AP effect can then be expressed in terms of a simple algebraic expression
\begin{equation}
    P^{\ell'}(k'; z) = \sum_m k'^{\tau_m} \underbrace{\sum_{\ell} \mathcal{P}^{\ell}_m(z) \mathcal{T}^{\ell \ell'}_m(z)}_{\equiv \mathcal{P}^{\ell'}_m(z)}.
    \label{eq:APEffectPowerSpectrumDecomposition}
\end{equation}
The transformation matrix of the AP effect, $\mathcal{T}^{\ell \ell'}_m(z)$, is defined as
\begin{equation}
    \mathcal{T}^{\ell \ell'}_m(z) = \frac{2\ell' + 1}{2} \frac{h(z)}{d^2(z)} \int_{-1}^{1} d\mu' \alpha^{\tau_m}(\mu'; z) \mathcal{L}_{\ell}\bigg(\frac{h(z)}{\alpha(\mu'; z)} \mu'\bigg) \mathcal{L}_{\ell'}(\mu') \, .
    \label{eq:APEffectTransformationMatrixIntegral}
\end{equation}

An analytical expression for this matrix can be derived by expressing the Legendre polynomials in terms of their Taylor polynomial,
\begin{equation}
    \mathcal{L}_{\ell}(x) = \sum_{n = 0}^{\ell} \kappa^{\ell}_n x^n \, ,
    \label{eq:APEffectLegendreTaylor}
\end{equation}
where\footnote{In particular, $\kappa^{\ell}_n = 0$ for $n > \ell$ or for odd $n + \ell$. Otherwise, $\kappa^{\ell}_n = (-1)^{(\ell+n) / 2} \frac{(\ell + n)!}{n! ((\ell+n)/2)! ((\ell-n)/2)!}$, as can be derived from eq. 22.3.8 in~\cite{AbramowitzStegun}.} $\kappa^{\ell}_n = \mathcal{L}_{\ell}^{(n)}(0) / n!$ with $\mathcal{L}_{\ell}^{(n)}(x) = d^n \mathcal{L}_{\ell}(x)/dx^n$, such that
\begin{equation}
    \mathcal{T}^{\ell \ell'}_m(z) = \frac{2\ell' + 1}{2} \frac{h(z)}{d^{2 + \tau_m}(z)} \sum_{n = 0}^{\ell} \sum_{n' = 0}^{\ell'} \kappa^{\ell}_n \kappa^{\ell'}_{n'} \big(h(z) d(z)\big)^n \int_{-1}^{1} d\mu' \frac{\mu'^{n+n'}}{\big(1 - \eta(z) \mu'^2\big)^{\frac{n - \tau_m}{2}}}\,.
    \label{eq:APEffectTransformationMatrix}
\end{equation}
The integral in this expression can now be solved analytically. First, note that when $n$ and $n'$ are of different parity, the integral vanishes. I.e., even though odd $\ell$s can be present due to relativistic effects~\cite{Gaztanaga:2015jrs,Guandalin:2020snp,Beutler:2020evf}, the AP effect does not introduce any cross-talk between multipoles with different parity. Furthermore, since $\kappa^{\ell}_n = 0$ for $n$ and $\ell$ of different parity, this directly implies 
$\mathcal{T}^{\ell \ell'}_m(z) = 0$ 
for $\ell$ and $\ell'$ of different parity. In other words, power does not leak between multipoles of different parity. 

In the  non-trivial case where $\ell$ and $\ell'$ are of the same parity, such that $n + n'$ is always even, the integral in~\eqref{eq:APEffectTransformationMatrix} is instead given by eq.~\eqref{eq:AppInt-1}, such that the transformation matrix can finally be written as
\begin{equation}\label{eq:Tellellprime}
    \mathcal{T}^{\ell \ell'}_m(z) \,=\, \left\{
    \begin{array}{ll}
        \begin{aligned}
        (2\ell' + &1) \frac{h(z)}{d^{2+\tau_m}(z)} \sum_{n = 0}^{\ell} \sum_{n' = 0}^{\ell'} \frac{\kappa^{\ell}_{n} \kappa^{\ell'}_{n'}}{(n + n') + 1} \big(h(z) d(z)\big)^{n} \\
        &\times {}_2F_1\left(\frac{n - \tau_m}{2}, \frac{(n + n') + 1}{2}; \frac{(n + n') + 3}{2}; \eta(z)\right)
        \end{aligned} 
    & \quad \!\textrm{($\ell+\ell'$ even),\!} \\
       0 &  \quad \!\textrm{($\ell+\ell'$ odd),}
    \end{array} \right.
\end{equation}
where ${}_2F_1$ is the Gaussian hypergeometric function.\footnote{For the implementation of the hypergeometric function in the code, see Appendix~\ref{app:HyperGauss}.}

\subsection{The FFTLog Decomposition}\label{sec:fft}

Generally, the power spectrum does not admit the shape of eq.~\eqref{eq:APEffectPowerSpectrumMuPowerLaw} or eq.~\eqref{eq:APEffectPowerSpectrumMultipolePowerLaw} for the method of the previous section to be applicable for the AP effect. However, with the emergence of the FFTLog~\cite{Hamilton:1999uv} as a tool for the efficient treatment of the one-loop integrals of the power spectrum in the EFTofLSS~\cite{McEwen:2016fjn, Simonovic:2017mhp} (also see Section~\ref{sec:spt}), it can also be used to decompose a generic power spectrum shape into power-laws for the purposes of Section~\ref{sec:ap}.

The starting point of the FFTLog is a sample of a generic function, $f(x)$, on an evenly spaced logarithmic grid that may additionally be modulated with a power-law, $\sim x^{-\nu}$, effectively introducing a \textit{bias} to the original signal, whose primary role will be the convergence of the one-loop integrals~\cite{McEwen:2016fjn, Simonovic:2017mhp} when considering $f = P^{(0)}$, the linear power spectrum,
\begin{equation}
    \bigg\{f(x_0) \, x_0^{-\nu}, \dots , f(x_{N-1}) \, x_{N-1}^{-\nu}\bigg\} \, ,
    \label{eq:PkSample}
\end{equation}
where $N$ is the size of the sample, $\log x_n = \log x_{\min} + n \Delta \log x$, and $\Delta \log x = 1/(N - 1)$ $\log x_{\max} / x_{\min}$. This sample is then analysed with a Fast Fourier Transform (FFT), for which the following convention is used
\begin{equation}
    c_m[f] = \sum_{n = 0}^{N - 1} f(x_n) \exp\bigg[ -2\pi i \frac{n m}{N} - \nu \log x_n \bigg] \equiv x_{\min}^{\eta_m} \sum_{n = 0}^{N - 1} f(x_n) \, x_n^{2 \nu_m} \, ,
    \label{eq:FFTLogDFT}
\end{equation}
where $\nu_m$ and $\eta_m$ are defined through:
\begin{equation}
    \nu_m = - \frac{1}{2} \bigg[\nu + 2\pi i \frac{m}{\log k_{\max}/k_{\min}} \frac{N-1}{N}\bigg] \equiv -\frac{1}{2} \bigg[\nu + \eta_m\bigg] \, .
    \label{eq:FFTLogExponentPowerLaw}
\end{equation}

Inverting the discrete Fourier transform (DFT) of eq.~\eqref{eq:FFTLogDFT}, the function $f$ can therefore be reconstructed in terms of power-laws as
\begin{equation}
    f(x) = \sum_{m = -\lfloor N/2 \rfloor}^{\lfloor N/2 \rfloor} \tilde{c}_m[f] \, x^{-2\nu_m} \equiv \sum_m \tilde{c}_m[f] \, x^{-2\nu_m} \, ,
    \label{eq:PkDecomposition}
\end{equation}
where $\lfloor \cdot \rfloor$ is the floor operator and the coefficients $\tilde{c}_m$ are given by\footnote{Note that $c^{*}_m = c_{N - m}$, such that $c_0 \in \mathbb{R}$ and, if $N$ is even, also $c_{N/2} \in \mathbb{R}$ (hence the factor of $1/2$ when $|m| = N/2$).}
\begin{equation}
    \tilde{c}_m[f] = \frac{x_{\min}^{-\eta_m}}{N} \left\{
    \begin{array}{cc}
        W_m \, c_m[f] & 0 \leq m < N/2 \\
        W_{-m} \, c^*_{-m}[f] & - N/2 < m < 0 \\
        \frac{1}{2} W_{N/2} \, c_{N/2}[f] & |m| = N/2
    \end{array}
    \right..
\end{equation}
Here, $c^*_{m}$ is the complex conjugate of $c_{m}$, and $W_m$ is an additional window function that is used to reduce ringing around sharp edges of the input signal, also known as Gibbs phenomenon. By default,\footnote{Additionally, the classic top-hat window function, the Lanczos window function or the Planck window function of~\cite{McKechan:2010kp} are available in the code.} we use the window function of~\cite{McEwen:2016fjn},
\begin{equation}
    W(x) =
    \left\{
    \begin{array}{cc}
        1 & x < x_0 \\
        \varphi(x) - \frac{1}{2 \pi} \sin\big(2\pi \, \varphi(x)\big) & x_0 \leq x \leq x_1 \\
        0 & x > x_1
    \end{array}\right. ,
    \label{eq:WindowFunction}
\end{equation}
where $\varphi(x) = (x_1 - x) / (x_1 - x_0)$ for $x_0$ and $x_1$ the bounds of the windowed region.

Note that even if the function $f$ does not exhibit discontinuities, the periodicity of the DFT~\eqref{eq:FFTLogDFT} in general introduces discontinuities in the boundaries, and the reconstructed signal~\eqref{eq:PkDecomposition} will thus be affected by ringing  (see also~\cite{McEwen:2016fjn} for a discussion). A secondary role of the bias $\nu$ is its tilting effect on the initial signal, which can mitigate such edge effects by closing the gap between $f(x_n)$ and $f(x_{n + N})$.

\subsection{Standard Perturbation Theory}\label{sec:spt}

In standard perturbation theory \cite{Peebles:1980yev,Fry:1983cj,Goroff:1986ep,Scoccimarro:1995if} (see \cite{Bernardeau:2001qr} for a review), the power spectrum attains the following form up until one-loop order
\begin{equation}
    P_{\textrm{SPT}}\big(\vec{k}; z\big) = P_{\textrm{tree,SPT}}\big(\vec{k}; z\big) + P_{\textrm{1-loop,SPT}}\big(\vec{k}; z\big) \, ,
\end{equation}
with the tree-level and one-loop order contributions given by
\begin{subequations}
    \begin{equation}
        P_{\textrm{tree,SPT}}\big(\vec{k}; z\big) = \Big[Z^{(s)}_1\big(\vec{k}; z\big)\Big]^2 P^{(0)}(k; z) \, ,
        \label{eq:Ptree}
    \end{equation}
    \begin{equation}
        \begin{aligned}
            P_{\textrm{1-loop,SPT}}\big(\vec{k}; z\big) &= 2 \int_{\vec{q}} \Big[Z^{(s)}_2\Big(\vec{q}, \vec{k}-\vec{q}; z\Big)\Big]^2 P^{(0)}(q; z) P^{(0)}\big(|\vec{k} - \vec{q}|; z\big) \\
            &+ 6 \, Z^{(s)}_1\big(\vec{k}; z\big) P^{(0)}(k; z) \int_{\vec{q}} Z^{(s)}_3\Big(\vec{q}, -\vec{q}, \vec{k}; z\Big) P^{(0)}(q; z) \, ,
        \end{aligned}
        \label{eq:P1loop}
    \end{equation}
\end{subequations}
where $\int_{\vec{q}} = \int \frac{dq^3}{(2\pi)^3}$. In these expressions $P^{(0)}(k; z) \equiv D^2_{+}(z) P^{(0)}(k)$ is the linear power spectrum at redshift $z$, and $D_{+}(z)$ is the linear growth factor. The explicit parametrisations of the kernels of biased tracers in redshift space, $Z^{(s)}_n$, are given in Appendix~\ref{app:Model}.

To compute the integrals of the one-loop part with the FFTLog approach, following~\cite{Simonovic:2017mhp}, the following decompositions are performed\footnote{The explicit coefficients $\mathcal{Z}^{22}_{ij}$ and $\mathcal{Z}^{13}_{ij}$ can be found in a Jupyter notebook provided with the code.}
\begin{subequations}
    \begin{equation}
        2 \bigg\langle \Big[Z^{(s)}_2\big(\vec{q}, \vec{k} - \vec{q}; z\big)\Big]^2 \bigg\rangle_{\phi_q} = \sum_{i, j} \qbar^{2i} \big|\hat{k} - \vec{\qbar}\big|^{2j} \mathcal{Z}^{22}_{ij}(\mu; z) \, ,
        \label{eq:Z2KernelDecomposition}
    \end{equation}
    \begin{equation}
        6 Z^{(s)}_1\big(\vec{k}; z\big) \bigg\langle Z^{(s)}_3\big(\vec{k}, \vec{q}, -\vec{q}; z\big) \bigg\rangle_{\phi_q} = \sum_{i, j} \qbar^{2i} \big|\hat{k} - \vec{\qbar}\big|^{2j} \mathcal{Z}^{13}_{ij}(\mu; z) \, ,
        \label{eq:Z3KernelDecomposition}
    \end{equation}
\end{subequations}
where $\qbar = q / k$ and $\langle \cdot \rangle_{\phi_q} = \frac{1}{2\pi} \int^{2\pi}_0 (\cdot) d\phi_q$ is the average over the azimuthal angle of the integration variable $\vec{q}$, if the coordinate system is chosen such that $\hat{k} = (0, 0, 1)$, $\hat{s} = (\sqrt{1 - \mu^2}, 0, \mu)$, and $\hat{q} = (\cos \phi_q \sin \theta_q, \sin \phi_q \sin \theta_q, \cos \theta_q)$ with $\cos \theta_q = \hat{k} \cdot \hat{q}$.

With this, as well as with the FFTLog decomposition of the linear power spectrum in the form of eq.~\eqref{eq:PkDecomposition}, the one-loop integrals of eq.~\eqref{eq:P1loop} become analytically solvable,
\begin{equation}
    P_{\textrm{1-loop,SPT}}\big(k, \mu; z\big) = k^3 \sum_{m_1, m_2} \mathcal{P}^{\textrm{1-loop,SPT}}_{m_1 m_2}(\mu; z) \, k^{-2(\nu_{m_1} + \nu_{m_2})} \, ,
    \label{eq:P1loopFFTLogPre}
\end{equation}
where
\begin{equation}
    \begin{aligned}
        \mathcal{P}^{\textrm{1-loop,SPT}}_{m_1 m_2}(\mu; z) = \tilde{c}_{m_1}\big[P^{(0)}\big](z) \, \tilde{c}_{m_2}\big[P^{(0)}\big](z) \sum_{i_1, i_2} \Big[&\mathcal{Z}^{22}_{i_1 i_2}(\mu; z) \, I(\nu_{m_1} - i_1, \nu_{m_2} - i_2) \\
        & + \mathcal{Z}^{13}_{i_1 i_2}(\mu; z) \, I(\nu_{m_1} - i_1, -i_2) \Big] \, ,
    \end{aligned}
    \label{eq:P1loopKernel}
\end{equation}
with the \textit{one-loop master integral} 
\begin{equation}
    I(\nu_1, \nu_2) = \int_{\vec{\qbar}} \frac{1}{\qbar^{2 \nu_1}} \frac{1}{\big|\hat{k} - \vec{\qbar}\big|^{2 \nu_2}} = \frac{1}{8 \pi^{3/2}} \frac{\Gamma(3/2 - \nu_1) \Gamma(3/2 - \nu_2) \Gamma(\nu_{12} - 3/2)}{\Gamma(\nu_1) \Gamma(\nu_2) \Gamma(3 - \nu_{12})} \, ,
    \label{eq:P1loopMasterIntegral}
\end{equation}
where $\nu_{12} = \nu_1 + \nu_2$. Expression~\eqref{eq:P1loopFFTLogPre} was first derived in~\cite{Simonovic:2017mhp}, and has since been implemented in codes such as \texttt{PyBird}~\cite{DAmico:2020kxu} and \texttt{CLASS-PT}~\cite{Chudaykin:2020aoj} for an efficient computation of the one-loop integrals.

A final improvement can be made by using the definition of $\nu_m$~\eqref{eq:FFTLogExponentPowerLaw}. In fact, using $-2(\nu_{m_1} + \nu_{m_2}) = 2\nu + \eta_{m_1 + m_2}$, and by collecting all terms with $m = m_1 + m_2$, which amounts to summing all elements along a given antidiagonal 
of the matrix $P^{\textrm{1-loop,SPT}}_{m_1 m_2}(\mu)$,
\begin{equation}
    \mathcal{P}^{\textrm{1-loop,SPT}}_{m}(\mu; z) \equiv \sum_{m_1+m_2 = m} \mathcal{P}^{\textrm{1-loop,SPT}}_{m_1 m_2}(\mu; z) \, ,
    \label{eq:P1loopDecompositionKernel}
\end{equation}
the original $\mathcal{O}(N^2)$ complexity of the matrix multiplication in eq.~\eqref{eq:P1loopFFTLogPre} is instead turned into the far more efficient $\mathcal{O}(N)$ complexity of the scalar product,
\begin{equation}
    P_{\textrm{1-loop,SPT}}\big(k, \mu; z\big) = k^{3 + 2\nu} \sum_{m} \mathcal{P}^{\textrm{1-loop,SPT}}_{m}(\mu; z) \, k^{\eta_m} \, .
    \label{eq:P1loopDecomposition}
\end{equation}
The additional cost is a sum over twice as many modes as before, namely from $m = -2 \lfloor N/2 \rfloor$ to $m = 2 \lfloor N/2 \rfloor$.

Note that this expression for the one-loop power spectrum already follows a power series, such that it can be directly used with the method of Section~\ref{sec:ap} for the AP effect. On the other hand, also the tree-level power spectrum can be expressed in terms of a power series by decomposing $P^{(0)}(k)$ in an FFTLog, 
\begin{equation}
    P_{\textrm{tree,SPT}}\big(k, \mu; z\big) = k^{\nu} \sum_m \mathcal{P}^{\textrm{tree,SPT}}_{m}(\mu; z) \, k^{\eta_m}
    \label{eq:PtreeDecomposition}
\end{equation}
with
\begin{equation}
    \mathcal{P}^{\textrm{tree,SPT}}_{m}(\mu; z) = \Big[Z^{(s)}_1(\mu; z)\Big]^2 \tilde{c}_m\big[P^{(0)}\big](z) \, .
\end{equation}

\subsection{UV Counterterms}\label{sec:uv}

In the EFTofLSS, the SPT power spectrum of the previous subsection needs to be amended with UV counterterms to cancel small-scale, non-linear contributions from the latter one-loop integral of eq.~\eqref{eq:P1loop}, the so-called $P_{13}$ contribution. Following~\cite{Perko:2016puo}, we adopt the following parametrisation of the UV counterterms
\begin{equation}
    P_{\textrm{1-loop,UV}}\big(k, \mu; z\big) = -2 Z^{(s)}_1(\mu; z) \Big(c_0(z) + c_2(z) f(z) \mu^2 + c_4(z) f^2(z) \mu^4\Big) k^2 P^{(0)}(k; z)\,.
    \label{eq:P1loopUV}
\end{equation}

Following the same argument for decomposing the tree-level power spectrum with the FFTLog into a power series for an efficient application of the AP effect with the method of Section~\ref{sec:ap}, also this contribution is decomposed in an FFTLog,
\begin{equation}
    P_{\textrm{1-loop,UV}}\big(k, \mu; z\big) = k^{2+\nu} \sum_m \mathcal{P}^{\textrm{1-loop,UV}}_{m}(\mu; z) \, k^{\eta_m} \, ,
    \label{eq:P1loopUVDecomposition}
\end{equation}
where
\begin{equation}
    \mathcal{P}^{\textrm{1loop,UV}}_{m}(\mu; z) = -2 Z^{(s)}_1(\mu; z) \Big(c_0(z) + c_2(z) f(z) \mu^2 + c_4(z) f^2(z) \mu^4\Big) \tilde{c}_m\big[P^{(0)}\big](z)\,.
\end{equation}

Additionally, we include the next-to-leading order (NLO) term of~\cite{Chudaykin:2020aoj, Ivanov:2019pdj},
\begin{equation}
    P^{\textrm{NLO}}_{\textrm{UV}}(k, \mu; z) = -\tilde{c}(z) \Big[Z^{(s)}_1(\mu; z)\Big]^2 \mu^4 f^4(z) k^4 P^{(0)}(k; z),
\end{equation}
as a proxy for higher order contributions from the fingers-of-God (FOG) effect. Although we also put this term subject to a similar FFTLog decomposition as eq.~\eqref{eq:P1loopUV}, which we omit to write down explicitly, we exclude it from the IR resummation in Section~\ref{sec:ir}, as we restrict ourselves to purely tree-level and one-loop order calculations.

\subsection{Stochasticity}\label{sec:stoch}

So far, only the deterministic part of the power spectrum has been considered. Stochastic contributions from the highly non-linear process of halo formation have to be added separately. Following~\cite{Perko:2016puo}, the stochastic part of the power spectrum up to one-loop order for biased tracers in redshift space is modelled as 
\begin{equation}
    P_{\textrm{stoch}}\big(k, \mu; z\big) = \frac{1}{\bar{n}} \left[c_{\epsilon 0}(z) + \Big(c_{\epsilon 1}(z) + c_{\epsilon 2}(z) f(z) \mu^2\Big) \frac{k^2}{k_M^2}\right],
    \label{eq:Stochasticity}
\end{equation}
where $k_M$ is the scale of halos and $\bar{n}$ the average number density of halos. Note that this equation is already of the desired shape of eq.~\eqref{eq:APEffectPowerSpectrumMuPowerLaw} for the efficient application of the AP effect, and thus needs no further modification.

\subsection{IR Resummation} \label{sec:ir}

Since Eulerian perturbation theory suffers from an intrinsic improper handling of IR displacements, leading to the broadening of the BAO peak, the so-called IR resummation was developed in~\cite{Lewandowski:2015ziq, Senatore:2014vja, Senatore:2014via} among others. By resumming linear displacements in the Lagrangian picture, the shortcoming can be amended. We write
\begin{equation}
    P^{\ell}(k; z)\Big\rvert_{\mathcal{N}} = 4 \pi \sum_{j = 0}^{\mathcal{N}} \sum_{\ell'} (-i)^{\ell'} \int^{\infty}_0 dq \, q^2 Q^{\ell \ell'}_{||\mathcal{N}-j}(k, q; z) \xi^{\ell'}(q; z)_j \, ,
    \label{eq:IRResumPowerSpectrum}
\end{equation}
where $P^{\ell}(k;z)\rvert_{\mathcal{N}}$  is the resummed power spectrum up to the $\mathcal{N}$'th loop order, and\footnote{Note that, in this section, the variable \(q\) has units of length.}
\begin{equation}
    \xi^{\ell}(q; z)_j = 4 \pi i^{\ell} \int_0^{\infty} \frac{dk k^2}{(2\pi)^3} P^{\ell}(k; z)_j \, j_{\ell}(kq) 
    \label{eq:IRResumCorrFunction}
\end{equation}
is the $j$-th loop piece of the (2-point) correlation function in Eulerian perturbation theory, related to the $j$-th loop piece of the power spectrum, $P^{\ell}(k;z)_j$, and $j_{\ell}(x)$ is the spherical Bessel function. Next, the functions $Q^{\ell \ell'}_{||\mathcal{N}-j}(k, q;z)$ are defined as
\begin{equation}
    Q^{\ell \ell'}_{||\mathcal{N}-j}(k, q; z) = \frac{i^{\ell'}}{4\pi} \frac{2\ell + 1}{2} \int^{1}_{-1} d\mu \int d^2\hat{q} \, e^{-i \vec{k} \cdot \vec{q}} F_{||\mathcal{N}-j}\big(\vec{k}, \vec{q}; z\big) \mathcal{L}_{\ell}(\mu) \mathcal{L}_{\ell'}(\mu_q) \, ,
    \label{eq:IRResumKernelQ}
\end{equation}
with $\mu_q = \hat{q} \cdot \hat{s}$ being the cosine of the angle of $\vec{q}$ w.r.t. the LOS $\hat{s}$, and
\begin{subequations}
    \begin{align}
        F_{||\mathcal{N}-j}\big(\vec{k}, \vec{q}; z\big) & = \,K_0\big(\vec{k}, \vec{q}; z\big) \bigg[K_0^{-1}\big(\vec{k}, \vec{q}; z\big)\Big\rvert\Big\rvert_{\mathcal{N}-j}\bigg] \, ,
        \label{eq:IRResumKernelF} \\
        K_0\big(\vec{k}, \vec{q}; z\big) & = \, \exp\bigg[-\frac{1}{2} k^2 \Xi_0(q; z)\bigg] \exp\bigg[-\frac{1}{2} k^2 \Xi_0(q; z) \mu^2 f(z) \big(2+f(z)\big)\bigg] \nonumber \\
        & \times \; \exp\bigg[-\frac{1}{2} k^2 \Xi_2(q; z)\big(\hat{k} \cdot \hat{q} + f(z)\mu\mu_q\big)^2\bigg] \, ,
        \label{eq:IRResumKernelK}
    \end{align}
\end{subequations}
where $K_0^{-1}||_{\mathcal{N}-j}$ means to expand $1/K_0$ to order $\mathcal{N}-j$ in Eulerian perturbation theory. Lastly, using the notation of~\cite{DAmico:2020kxu}, the objects $\Xi_0(q; z)$ and $\Xi_2(q; z)$ are defined as\footnote{In earlier works, cf.~\cite{Lewandowski:2015ziq, Senatore:2014vja, Senatore:2014via}, the notation $X(q)$ and $Y(q)$ was used for $\Xi_0(q)$ and $\Xi_2(q)$, respectively.}
\begin{subequations}
    \begin{align}
        \Xi_0(q; z) & \,=\, \frac{2}{3} \int^{\infty}_0 \frac{dk}{2\pi^2} P^{(0)}(k; z) \exp\left[-\frac{k^2}{\Lambda^2_{\textrm{IR}}}\right] \Big(1 - j_0(kq) - j_2(kq)\Big) \, ,
        \label{eq:IRFilterXi0} \\
        \Xi_2(q; z) &\,=\, 2 \int^{\infty}_0 \frac{dk}{2\pi^2} P^{(0)}(k; z) \exp\left[-\frac{k^2}{\Lambda^2_{\textrm{IR}}}\right] j_2(kq) \, ,
        \label{eq:IRFilterXi2}
    \end{align}
\end{subequations}
and act as filters for the correlation function, while the scale $\Lambda_{\textrm{IR}}$ controls the amount of IR displacements to include in the resummation (see also~\cite{Lewandowski:2015ziq}). Following~\cite{DAmico:2020kxu}, we use $\Lambda_{\textrm{IR}} = 0.12 \, h \textrm{Mpc}^{-1}$ by default.

\subsubsection{The IR Filters}\label{sec:ir-filters}

The IR filters $\Xi_0(q)$ and $\Xi_2(q)$ are the main ingredients in the resummation of the IR displacement fields. It is therefore worthwhile to discuss their numerical computation and their functional shapes, especially in the IR and UV regimes.

Already from eqs.~\eqref{eq:IRFilterXi0} and~\eqref{eq:IRFilterXi2}, the IR limit $q \to \infty$ of the filters can be read off directly. Since $\lim_{x \to \infty} j_{n}(x) = 0$ for $n \in \mathbb{N}_0$, we have
\begin{subequations}
    \begin{equation}
        \Xi^{\infty}_0(z) \equiv \lim_{q \to \infty} \Xi_0(q) = \frac{2}{3} \int^{\infty}_0 \frac{dk}{2\pi^2} P^{(0)}(k; z) \exp\left[-\frac{k^2}{\Lambda^2_{\textrm{IR}}}\right] \, ,
        \label{eq:IRFilterXi0Inf}
    \end{equation}
    \begin{equation}
        \Xi^{\infty}_2(z) \equiv \lim_{q \to \infty} \Xi_2(q) = 0 \, .
        \label{eq:IRFilterXi2Inf}
    \end{equation}
\end{subequations}
To describe the UV behaviour of the filters, we note that for small $x$, $1-j_0(x)-j_2(x) = x^2/10 + {\cal O}(x^4)$ and $j_2(x) = x^2/15 + {\cal O}(x^4)$ and find
\begin{equation}
    \Xi_n(q; z) = \Xi^{(2)}_n(z) \, q^2 + \mathcal{O}\big(q^4\big) \quad (n = 0,2) \,,
    \label{eq:IRResumFiltersAsymptote}
\end{equation}
where
\begin{subequations}
    \begin{align}
        \Xi^{(2)}_0(z) & = \, \frac{1}{30 \pi^2} \int_0^{\infty} dk \, k^{2} P^{(0)}(k; z) \exp\bigg[-\frac{k^2}{\Lambda_{\textrm{IR}}^2}\bigg] \,,
        \label{eq:IRResumFilterXi0Deriv} \\
        \Xi^{(2)}_2(z) & = \,2\, \Xi^{(2)}_0(z) \,.
        \label{eq:IRResumFilterXi2Deriv}
    \end{align}
\end{subequations}

\begin{figure}[t]
    \centering
    \includegraphics[width=\textwidth]{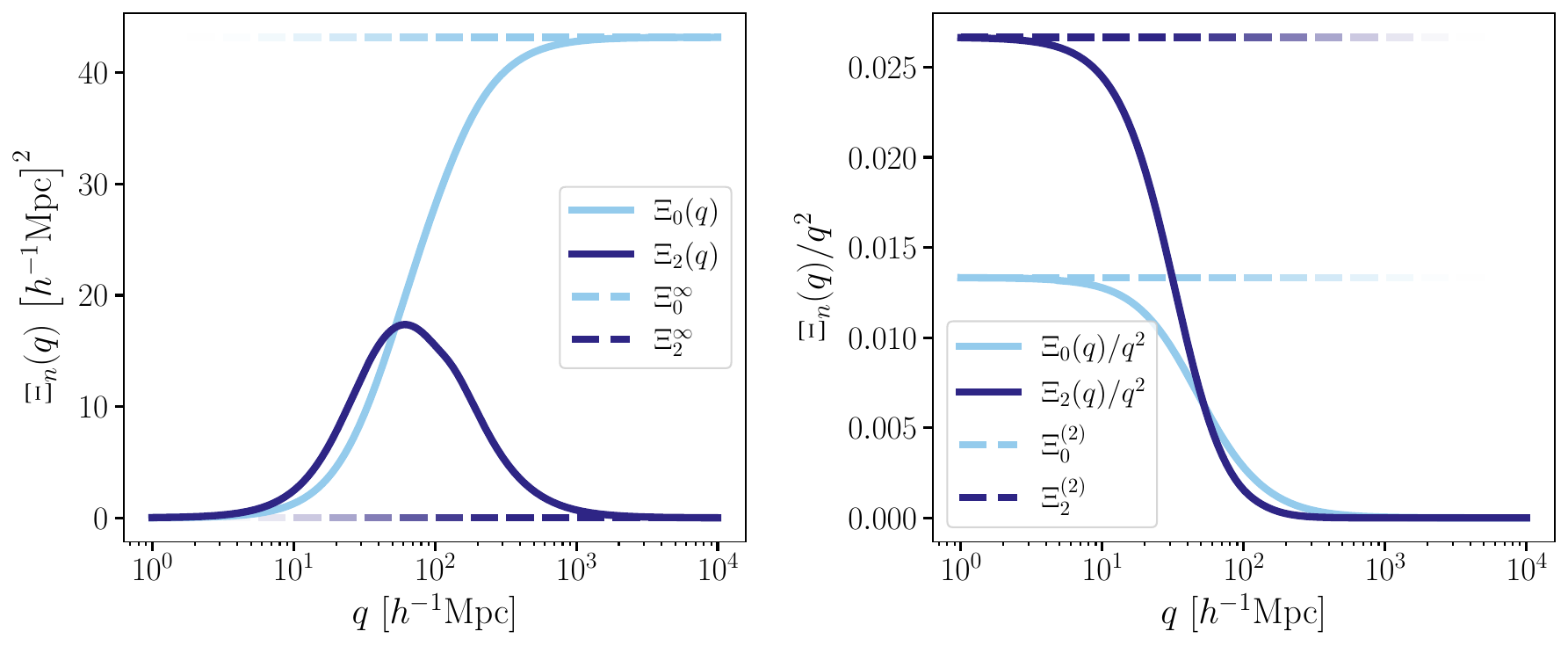}
    \caption{The IR filters $\Xi_0(q)$ and $\Xi_2(q)$ (left) with the asymptotes $\Xi^{\infty}_n = \lim_{q \to \infty} \Xi_n(q)$ and the rescaled filters $\Xi_0(q) / q^2$ and $\Xi_2(q) / q^2$ (right) with the asymptotes $\Xi^{(2)}_n = \lim_{q \to 0} \Xi_n(q) / q^2$. The plots were produced using $\Lambda_{\textrm{IR}} = 0.12 \, h \textrm{Mpc}^{-1}$ and the Planck-like $\Lambda$CDM cosmology described in Section~\ref{sec:comparison}.}
    \label{fig:ir-filters}
\end{figure}

As we show in the plot on the left in figure~\ref{fig:ir-filters},  $\Xi_0(q)$ acts as a high-pass filter, while $\Xi_2(q)$ is instead a band-pass filter centred around the BAO scale at $q \sim 100 \, h^{-1} \textrm{Mpc}$. On the other hand, by rescaling the filters with $1/q^2$, both filters turn into low-pass filters, as evident in the plot on the right in figure~\ref{fig:ir-filters}. In the following subsections, we will show the merit of the two sets of filters and their use in solving the integral in eq.~\eqref{eq:IRResumPowerSpectrum} also on non-linear scales.

As for the numerical computation of the filters $\Xi_0(q)$ and $\Xi_2(q)$ themselves, similarly as in \texttt{PyBird}, we make use of the FFTLog. For this, we first define
\begin{equation}
    \hat{\Xi}_n(q; z) = 2 \int^{\infty}_0 \frac{dk}{2\pi^2} P^{(0)}(k; z) \exp\left[-\frac{k^2}{\Lambda^2_{\textrm{IR}}}\right] j_n(kq) \quad (n = 0, 2) \, ,
    \label{eq:IRFilterHatXi}
\end{equation}
such that $\Xi_0(q) = \Xi^{\infty}_0 + (\hat{\Xi}_0(q) + \hat{\Xi}_2(q)) / 3$ and $\Xi_2(q) = \hat{\Xi}_2(q)$. Next, by performing an FFTLog of $P^{(0)}(k)$, eq.~\eqref{eq:IRFilterXi0Inf} becomes
\begin{align}
    \Xi^{\infty}_0(z) &= \frac{2}{3} \sum_{m} \tilde{c}_m\big[P^{(0)}\big](z) \int_0^{\infty} \frac{dk}{2 \pi^2} k^{-2\nu_m} \exp\left[-\frac{k^2}{\Lambda_{\textrm{IR}}^2}\right] \nonumber \\
    &= \frac{1}{3 \pi^2} \sum_{m} \tilde{c}_m\big[P^{(0)}\big](z) \, I_2(\Lambda_{\textrm{IR}}^{-2}, -2\nu_m)\,,
    \label{eq:IRFilterXi0InfFFTLog}
\end{align}
while eq.~\eqref{eq:IRResumFilterXi0Deriv} similarly becomes,
\begin{align}
    \Xi^{(2)}_0(z) &= \frac{1}{30 \pi^2} \sum_{m} \tilde{c}_m\big[P^{(0)}\big](z) \int_0^{\infty} dk \, k^{2-2\nu_m} \exp\left[-\frac{k^2}{\Lambda_{\textrm{IR}}^2}\right] \nonumber \\
    &= \frac{1}{30 \pi^2} \sum_{m} \tilde{c}_m\big[P^{(0)}\big](z) \, I_2(\Lambda_{\textrm{IR}}^{-2}, 2-2\nu_m)\,,
    \label{eq:IRResumFilterXi0DerivFFTLog}
\end{align}
with $I_2$ the solution of the integral as defined in eq.~\eqref{eq:AppIntExpPow}. Lastly, eq.~\eqref{eq:IRFilterHatXi} is solved by performing an FFTLog of $\tilde{P}^{(0)}(k) \equiv P^{(0)}(k) \exp[-k^2/\Lambda_{\textrm{IR}}^2]$,
\begin{align}
    \hat{\Xi}_n(q; z) &= 2 \sum_{m} \tilde{c}_m\big[\tilde{P}^{(0)}\big](z) \int_0^{\infty} \frac{dk}{2 \pi^2} k^{-2\nu_m} j_n(kq) \nonumber \\
    &= \frac{1}{\pi^2} \sum_{m} \tilde{c}_m\big[\tilde{P}^{(0)}\big](z) \, I_3(n, -2\nu_m) \,  q^{2\nu_m-1}\,,
    \label{eq:IRFilterHatXiFFTLog}
\end{align}
where $I_3$ is given by eq.~\eqref{eq:AppIntJnPow}.

\subsubsection{Direct Method}\label{sec:ir-direct}

For an efficient implementation of the IR resummation~\eqref{eq:IRResumPowerSpectrum} into the FFTLog framework, the following perturbative expansion was introduced in~\cite{DAmico:2020kxu} and implemented in \texttt{PyBird},
\begin{align}
    P^{\ell}(k; z)\Big\rvert_{\mathcal{N}} = P^{\ell}(k; z)\Big\rvert\Big\rvert_{\mathcal{N}} + 4 \pi & \sum_{j = 0}^{\mathcal{N}} \sum_{n = 1}^{N} k^{2n} \sum_{n_0 = N_{0i}(n)}^{N_{0f}(n)} \sum_{\ell'} (-i)^{\ell'} \sum_{\alpha = A_i(\ell', n_0)}^{A_f(\ell', n_0)} \mathcal{Q}^{\ell \ell'}_{n n_0 \alpha || \mathcal{N}-j}(z) \nonumber \\
    &\times \int^{\infty}_0 dq \, q^2 \, \Xi_0^{n_0}(q; z) \Xi_2^{n-n_0}(q; z) \xi^{\ell'}(q; z)_j \, j_{\alpha}(kq) \,, \label{eq:IRResumPowerSpectrumDirect}
\end{align}
which follows from eq.~\eqref{eq:IRResumPowerSpectrum} by directly expanding each exponential term in eq.~\eqref{eq:IRResumKernelK} individually to orders $N_{0a}$, $N_{0b}$, and $N_2$, respectively, giving the summation bounds $N = N_{0a} + N_{0b} + N_2$, $N_{0i}(n) = \max(0, n - N_2)$ and $N_{0f}(n) = \min(n, N_{0a} + N_{0b})$, as well as $A_i(\ell', n_0) = \max(0, \ell' - 2n_0)$ and $A_f(\ell', n_0) = \ell' + 2n_0$. Here, $P^{\ell}(k; z)\rvert\rvert_{\mathcal{N}}$ is the (unresummed) power spectrum up to the $\mathcal{N}$'th loop order, i.e., the power spectrum in Eulerian perturbation theory. The coefficients $\mathcal{Q}^{\ell \ell'}_{n n_0 \alpha || \mathcal{N}-j}(z)$ that have been introduced in~\cite{DAmico:2020kxu}, albeit without an explicit expression, can be written as
\begin{equation}
    \mathcal{Q}^{\ell \ell'}_{n n_0 \alpha ||\mathcal{N} - j}(z) = \frac{(-1)^n}{2^n} \sum_{\kappa = 0}^{\mathcal{N}-j} (-1)^{\kappa} \sideset{}{'}\sum_{\substack{\kappa_{0a}, \kappa_{0b}, \kappa_2 \\ \sum_i \kappa_i = \kappa}} \left.\left(\prod_{i \in \{0a, 0b, 2\}} \, \sideset{}{'}\sum_{n_i = \kappa_i}^{N_i}\right)\right\rvert_{\begin{subarray}{l} n_{0a} + n_{0b} = n_0 \\ n_2 = n - n_0 \end{subarray}} \mathcal{J}^{\ell \ell'}_{n_{0b} n_2\alpha}(z),
    \label{eq:IRResumCoefQDirect}
\end{equation}
where we introduced the notation $(\sum')_{i=i_0} = \sum_{i = i_0} \frac{1}{(i - i_0)!}$, and where $\mathcal{J}^{\ell \ell'}_{n_{0b} n_2 \alpha}(z)$ is given by
\begin{equation}
    \begin{aligned}
        \mathcal{J}^{\ell \ell'}_{n_{0b} n_2 \alpha}(z) &= \, \frac{2 \ell + 1}{2} (-1)^{n_2} \Big[f(z) \big(2+f(z)\big)\Big]^{n_{0b}} \\
        &\times\sum_{i_2 = 0}^{2 n_2} \binom{2 n_2}{i_2} f^{i_2}(z) \sum_{j_2 = -i_2}^{i_2} (-i)^{i_2 + j_2} \frac{2 (\ell' + j_2) + 1}{2} x^{i_2}_{\ell', \ell' + j_2} \, x^{2n_{0b} + i_2}_{\ell, \ell' + j_2} \, a^{2n_2 - i_2}_{\ell' + j_2, \alpha} \,,
    \end{aligned}
    \label{eq:IRResumCoefJ}
\end{equation}
with $x^{n}_{\ell \ell'}$ and $a^{n}_{\ell \ell'}$ being defined in eq.~\eqref{eq:AppIRResumDoubleLegendreIntegral} and eq.~\eqref{eq:AppIRResumBesselDerivativeCoef}, respectively. For a derivation, see Appendix~\ref{app:ir-direct}.

Following the same procedure as in \texttt{PyBird}, the remaining integral in eq.~\eqref{eq:IRResumPowerSpectrumDirect},
\begin{equation}
    I^{\ell'}_{n_0 n_2 \alpha}(k; z)_j \equiv \int^{\infty}_0 dq \, q^2 \, \Xi_0^{n_0}(q; z) \Xi_2^{n_2}(q; z) \xi^{\ell'}(q; z)_j j_{\alpha}(kq)\,,
    \label{eq:IRResumMasterIntegralDirect}
\end{equation}
is solved semi-analytically by performing an FFTLog of the filtered correlation function, $\Xi_0^{n_0}(q) \Xi_2^{n_2}(q) \xi^{\ell'}(q)_j$,
\begin{align}
    I^{\ell'}_{n_0 n_2 \alpha}(k; z)_j &= \sum_{m} \tilde{c}_m\Big[\Xi^{n_0}_0 \, \Xi^{n_2}_2 \, \xi^{\ell'}_j \Big](z) \int^{\infty}_0 dq \, q^{2-2\nu_m} j_{\alpha}(kq) \nonumber \\
    &=\sum_{m} \tilde{c}_m\Big[\Xi^{n_0}_0 \, \Xi^{n_2}_2 \, \xi^{\ell'}_j \Big](z) \, I_3(\alpha, 2 - 2\nu_m) \, k^{-3 + 2\nu_m},
    \label{eq:IRResumMasterIntegralDirectFFTLog}
\end{align}
with $I_3$ the integral solution as given in eq.~\eqref{eq:AppIntJnPow}.

By default, we use $N_{0a} = 5$, $N_{0b} = 6$, and $N_2 = 7$ for the bounds of the expansion. We have checked that higher orders are negligible to about $1 \%$ on scales of $k \lesssim 0.4 \, h\textrm{Mpc}^{-1}$ for the Planck-like cosmology described in Section~\ref{sec:comparison} at $z = 0$ (see also figure~\ref{fig:ir-resum}).

\subsubsection{Asymptotic Method}\label{sec:ir-asympt}

The method presented above unfortunately fails to converge on small enough scales. For $z=0$, the point at which the previous method fails to converge is around $k \gtrsim 0.45 \, h \textrm{Mpc}^{-1}$  (cf.~figure~\ref{fig:ir-resum}), with higher redshifts pushing this threshold to even higher values of $k$. Since the EFTofLSS at one-loop order is expected to be inaccurate on these scales~\cite{Carrasco:2012cv, Carrasco:2013mua}, one may be content with the direct method. However, not only could an extension to two-loop order in the EFTofLSS (e.g., by following~\cite{Simonovic:2017mhp}) potentially give access to these scales~\cite{Carrasco:2013mua,Konstandin:2019bay}, already the correlation function at one-loop order, viewed as an integral over the power spectrum, benefits from a smooth extension of the IR resummation into the non-linear regime, as we will show later explicitly. We therefore present such an extension already in this work.

The reason for the divergence is mostly attributed to the perturbative character of the direct method, stemming from the expansion of $K_0(\vec{k}, \vec{q})$~\eqref{eq:IRResumKernelK}. With larger values of $k$, the Bessel functions in eq.~\eqref{eq:IRResumPowerSpectrumDirect} shift the support to smaller values of $q \sim k^{-1}$. In turn, as more terms are needed in the expansion to achieve convergence, given the high-pass nature of $\Xi_0(q)$ and the band-pass nature of $\Xi_2(q)$ (cf.~figure~\ref{fig:ir-filters}), the filtered correlation function, $\Xi^{n_0}_0(q) \Xi^{n_2}_2(q) \xi^{\ell}(q)_j$, therefore extends over several orders of magnitude on the $q$ half-line. As a result, the FFTLog in eq.~\eqref{eq:IRResumMasterIntegralDirectFFTLog} fails to resolve the features at the scales of interest $q \sim k^{-1}$, leading to the divergence of the overall integral.

These issues can be resolved by using the rescaled filters
\begin{equation}
    \tilde{\Xi}_n(q; z) \equiv \frac{\Xi_n(q; z)}{q^2} \quad (n = 0,2)
\end{equation}
in the FFTLog instead, as they behave like low-pass filters (cf. figure~\ref{fig:ir-filters}) that place emphasis on the required small scale modes and filter out the unwanted large scale modes. This allows the FFTLog to once again resolve the required region of interest, and the integral~\eqref{eq:IRResumMasterIntegralDirectFFTLog} to converge.

However, due to the additional factors of $q^{2(n_0 + n_2)}$ that would appear in eq.~\eqref{eq:IRResumMasterIntegralDirectFFTLog} with the rescaled filters, the resulting integral $I_3(\alpha, 2 - 2\nu_m + 2(n_0 + n_2))$ then only converges for values of the FFTLog's bias of $\nu < -1 - 2(n_0 + n_2)$ (cf. eq.~\eqref{eq:AppIntJnPow}). Due to the tilting effect of the bias on the overall signal in the FFTLog, such large negative biases cause large jumps at the borders of the sampled region, effectively producing uncontrolled ringing. To circumvent this issue, we make use of the special form of eq.~\eqref{eq:IRResumKernelK} for $K_0(\vec{k}, \vec{q})$ and extract a Gaussian to suppress the high $q$ behaviour of the integral, allowing for the use of a moderate value for the bias,
\begin{align}
    K_0\big(\vec{k}, \vec{q}; z\big) & = \, \exp\bigg[-\frac{1}{2} k^2 q^2 \bigg\{\Xi^{(2)}_0(z)\big(1 + f(z)\big)^2 + \Xi^{(2)}_2(z) \big(1 + f(z)\big)^2\bigg\}\bigg] \label{eq:IRResumKernelKAsymptote} \\
    &\times \; \exp\bigg[-\frac{1}{2} k^2 q^2 \bigg\{\tilde{\Xi}_0(q; z) \Big(1 + f(z)\big(2+f(z)\big) \mu^2\Big) - \Xi^{(2)}_0(z) \big(1 + f(z)\big)^2\bigg\}\bigg] \nonumber \\
    & \times \; \exp\bigg[-\frac{1}{2} k^2 q^2 \bigg\{\tilde{\Xi}_2(q; z) \big(\hat{k} \cdot \hat{q} + f(z)\mu\mu_q\big)^2 - \Xi^{(2)}_2(z) \big(1 + f(z)\big)^2\bigg\}\bigg] \nonumber.
\end{align}

As we show in more detail in Appendix~\ref{app:ir-asympt}, the next step in deriving the coefficients of the asymptotic method of the IR resummation is to expand only the latter two exponential functions. For this, we  now briefly comment on the specific form of the extracted terms and their role in guaranteeing the convergence of the expansion.

\begin{figure}[t]
    \centering
    \includegraphics[width=\textwidth]{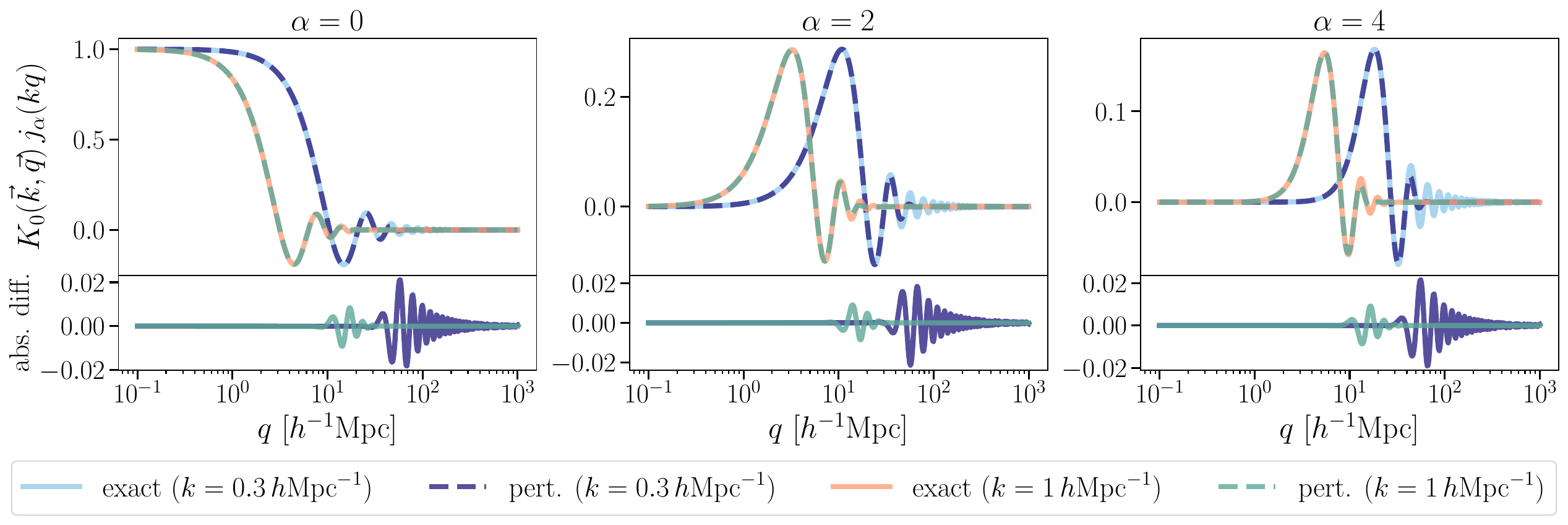}
    \caption{
    Comparison of the perturbed  with the full, exact,  $K_0\big(\vec{k}, \vec{q}\big) j_{\alpha}(kq)$ for $\alpha = 0, 2, 4$ in the worst-case scenario: $\hat{k} \cdot \hat{s} = 0$ and $\hat{k} \cdot \hat{q} = 0$. We consider the cases $k = 0.3 \, h \textrm{Mpc}^{-1}$ and $k = 1 \, h \textrm{Mpc}^{-1}$ for a Planck-like $\Lambda$CDM cosmology at $z = 0$ (cf. Section~\ref{sec:comparison}). The perturbed $K_0$ in eq.~\eqref{eq:IRResumKernelKAsymptote} has its second exponential expanded to $N_0 = 5$ and its third exponential expanded to $N_2 = 6$. 
    }
    \label{fig:ir-k0}
\end{figure}

First, notice that
\begin{equation}
    \max_{\hat{k}, \hat{q}} \Big(1 + f(z) \big(2 + f(z)\big) \mu^2\Big) = \big(1 + f(z)\big)^2 = \max_{\hat{k}, \hat{q}} \Big(\big(\hat{k}\cdot\hat{q} + f(z) \mu \mu_q\big)^2\Big).
\end{equation}
Furthermore, since $\tilde{\Xi}_n(q) \leq \Xi^{(2)}_n \; (\forall q \in \mathbb{R}^+)$ for $n = 0, 2$ (cf.~figure~\ref{fig:ir-filters}),\footnote{This result is not cosmology-dependent, but general. In fact, from Eqs.~\eqref{eq:IRFilterXi0} and~\eqref{eq:IRFilterXi2}, the conditions $\Xi_n(q) \leq \Xi^{(2)}_n q^2 \; (\forall q \in \mathbb{R}^+)$ for $n = 0, 2$ are equivalent to $1 \leq j_0(x) + j_2(x) + x^2/10 \; (\forall x \in \mathbb{R}^+)$ and $j_2(x) \leq x^2 / 15 \; (\forall x \in \mathbb{R}^+)$, respectively. One can verify that these inequalities are indeed true, e.g., via Mathematica.} the arguments of the latter two exponentials in eq.~\eqref{eq:IRResumKernelKAsymptote} are positive semi-definite for all $\vec{k}, \vec{q} \in \mathbb{R}^3$. This is important for the perturbative expansion, as $\exp(x) \leq \sum_{n = 0}^N x^n/n! \; (\forall N \in \mathbb{N}_0, \forall x \in \mathbb{R}^{+}_0)$, i.e., the exponential is never overestimated, but at worst only underestimated. This is not the case for negative arguments, given the resulting alternating power series.

As is evident from eq.~\eqref{eq:IRResumKernelKAsymptote}, the underestimation becomes most severe for $\mu = \hat{k} \cdot \hat{s} \to 0$ and $\hat{k} \cdot \hat{q} \to 0$ in combination with large values of $q$. However, since these larger values of $q$ are suppressed by the Bessel functions thanks to their decaying properties and their oscillatory nature, also the underestimation of $K_0(\vec{k}, \vec{q})$ is suppressed.

For instance, as we show qualitatively in figure~\ref{fig:ir-k0} for the Planck-like $\Lambda$CDM cosmology at $z = 0$ described in Section~\ref{sec:comparison}, in the worst-case scenario of $\mu = 0$ and $\hat{k} \cdot \hat{q} = 0$, the perturbative approach to eq.~\eqref{eq:IRResumKernelKAsymptote} correctly reproduces the exact results until well within the oscillatory region of the Bessel functions, and that already at $k = 0.3 \, h \textrm{Mpc}^{-1}$. For larger values of $k$, such as $k = 1 \, h \textrm{Mpc}^{-1}$, the agreement improves noticeably, as the Bessel functions decay more quickly.

Lastly, due to the presences of the Legendre polynomials, $\mathcal{L}_{\ell}(\mu)$ and $\mathcal{L}_{\ell'}(\mu_q)$ in eq.~\eqref{eq:IRResumKernelQ}, as well as the accompanying Legendre polynomial $\mathcal{L}_{\alpha}(\hat{k} \cdot \hat{q})$ of the Bessel functions,\footnote{These Legendre polynomials are hidden in the coefficients~\eqref{eq:IRResumCoefQDirect}, but basically enter the derivation in Appendix~\ref{app:ir-asympt} through Rayleigh's expansion $e^{-i \vec{k} \cdot \vec{q}} = \sum_{\alpha = 0}^{\infty} (-i)^{\alpha} (2 \alpha + 1) \mathcal{L}_{\alpha}(\hat{k} \cdot \hat{q}) j_{\alpha}(kq)$.} perpendicular angles are additionally penalised due to $|\mathcal{L}_n(x)| \leq |\mathcal{L}_n(\pm 1)|$ ($\forall n \in \mathbb{N}_0, \forall x \in [-1, 1]$). This further diminishes the effect of the underestimation of $K_0(\vec{k}, \vec{q})$ in the angular integrals, as straighter angles are weighted more strongly than perpendicular angles.

In any case, following the derivation of Appendix~\ref{app:ir-asympt}, the IR resummed power spectrum, eq.~\eqref{eq:IRResumPowerSpectrum}, then assumes the alternative form,
\begin{align}
    P^{\ell}(k; z)\Big\rvert_{\mathcal{N}} \, = \, & \, 4 \pi \sum_{j = 0}^{\mathcal{N}} \sum_{n = 0}^{N} k^{2n} \sum_{\tilde{n} = 0}^{n} \sum_{\tilde{n}_0 = N_{0i}(\tilde{n})}^{N_{0f}(\tilde{n})} \sum_{\ell'} (-i)^{\ell'} \sum_{\alpha = A_i(\ell', \tilde{n}_0)}^{A_f(\ell', \tilde{n}_0)} \tilde{\mathcal{Q}}^{\ell \ell'}_{n \tilde{n} \tilde{n}_0 \alpha || \mathcal{N}-j}(z) \label{eq:IRResumPowerSpectrumAsymptote} \\
    &\times \int^{\infty}_0 dq \, q^{2+2n} \exp\bigg[-k^2 q^2 \Xi^{(2)}(z)\bigg] \tilde{\Xi}^{\tilde{n}_0}_0(q; z) \tilde{\Xi}^{\tilde{n} - \tilde{n}_0}_2(q; z) \xi^{\ell'}(q; z)_j \, j_{\alpha}(kq) \,,\nonumber
\end{align}
where now $N = N_0 + N_2$, with $N_0$ and $N_2$ the respective orders to which the latter two exponentials of eq.~\eqref{eq:IRResumKernelKAsymptote} are expanded, and where $\Xi^{(2)}(z)$ has been defined as
\begin{equation}
    \Xi^{(2)}(z) = \frac{1}{2} \bigg[\Xi^{(2)}_0(z)\big(1 + f(z)\big)^2 + \Xi^{(2)}_2(z) \big(1 + f(z)\big)^2\bigg]\,.
\end{equation}
The coefficients $\tilde{\mathcal{Q}}^{\ell \ell'}_{n\tilde{n}\tilde{n}_0\alpha || \mathcal{N}-j}(z)$ have been introduced as
\begin{equation}
    \begin{aligned}
        \tilde{\mathcal{Q}}^{\ell \ell'}_{n \tilde{n} \tilde{n}_0 \alpha ||\mathcal{N} - j}(z) &= \frac{(-1)^{\tilde{n}}}{2^n} \sum_{\kappa = 0}^{\mathcal{N}-j} (-1)^{\kappa} \sideset{}{'}\sum_{\substack{\kappa_{0}, \kappa_2 \\ \sum_i \kappa_i = \kappa}} \left.\left(\prod_{i \in \{0,2\}} \, \sideset{}{'}\sum_{n_i = \kappa_i}^{N_i} \, \sideset{}{''}\sum_{m_i = \kappa_i}^{n_i}\right)\right\rvert_{\begin{subarray}{l} m_0 = \tilde{n}_0 \\ m_2 = \tilde{n} - \tilde{n}_0 \end{subarray}} \\
        &\times \Big[\Xi^{(2)}_0(z) \, \big(1 + f(z)\big)^2\Big]^{n_0 - m_0} \Big[\Xi^{(2)}_2(z) \, \big(1 + f(z)\big)^2\Big]^{n_2 - m_2} \, \tilde{\mathcal{J}}^{\ell \ell'}_{m_{0} m_2\alpha}(z)\,,
    \end{aligned}
    \label{eq:IRResumCoefQAsymptote}
\end{equation}
with the notation $(\sum'')_{i = i_0}^{i_1} = \sum_{i = i_0}^{i_1} \binom{i_1 - i_0}{i - i_0}$, and where
\begin{equation}
    \begin{aligned}
        \tilde{\mathcal{J}}^{\ell \ell'}_{m_0 m_2 \alpha}(z) = \, \frac{2 \ell + 1}{2} (-1)^{m_2} &\sum_{i_0 = 0}^{m_0} \binom{m_0}{i_0} \Big[f(z) \big(2+f(z)\big)\Big]^{i_0} \sum_{i_2 = 0}^{2 m_2} \binom{2 m_2}{i_2} f^{i_2}(z) \\
        \times &\sum_{j_2 = -i_2}^{i_2} (-i)^{j_2 + i_2} \frac{2 (\ell' + j_2) + 1}{2} x^{i_2}_{\ell', \ell' + j_2} \, x^{2i_0 + i_2}_{\ell, \ell' + j_2} \, a^{2m_2 - i_2}_{\ell' + j_2, \alpha} \,.
    \end{aligned}
    \label{eq:IRResumCoefJAsympt}
\end{equation}

Similarly to the direct method, the remaining integral in eq.~\eqref{eq:IRResumPowerSpectrumAsymptote},
\begin{equation}
    \tilde{I}^{\ell'}_{n \tilde{n}_0 \tilde{n}_2 \alpha}(k; z)_j \equiv \int^{\infty}_0 dq \, q^{2+2n} \exp\bigg[-k^2 q^2 \Xi^{(2)}(z)\bigg] \tilde{\Xi}_0^{\tilde{n}_0}(q; z) \tilde{\Xi}_2^{\tilde{n}_2}(q; z) \xi^{\ell'}(q; z)_j \, j_{\alpha}(kq),
    \label{eq:IRResumMasterIntegralAsymptote}
\end{equation}
can be solved semi-analytically by performing an FFTLog of the (rescaled) filtered correlation function, $\tilde{\Xi}_0^{\tilde{n}_0}(q) \tilde{\Xi}^{\tilde{n}_2}_2(q) \xi^{\ell'}_j(q)$,
 \begin{align}
    \tilde{I}^{\ell'}_{n \tilde{n}_0 \tilde{n}_2 \alpha}(k; z)_j &= \sum_{m} \tilde{c}_m\Big[\tilde{\Xi}_0^{\tilde{n}_0} \, \tilde{\Xi}^{\tilde{n}_2}_2 \, \xi^{\ell'}_j \Big](z) \int^{\infty}_0 dq \, q^{2+2n-2\nu_m} \exp\Big[-k^2 q^2 \Xi^{(2)}(z)\Big] j_{\alpha}(kq) \nonumber \\
    &=\sum_{m} \tilde{c}_m\Big[\tilde{\Xi}^{\tilde{n}_0}_0 \, \tilde{\Xi}^{\tilde{n}_2}_2 \, \xi^{\ell'}_j \Big](z) \, I_4\big(\Xi^{(2)}(z), \alpha, 2 + 2n - 2\nu_m\big) \, k^{-3 + 2\nu_m - 2n},
    \label{eq:IRResumMasterIntegralAsymptoteFFTLog}
\end{align}
with $I_4$ the integral solution as given in eq.~\eqref{eq:AppIntJnExpPow}. Inserting this expression into eq.~\eqref{eq:IRResumPowerSpectrumAsymptote} finally gives the desired asymptotic expansion, as the remaining $k$-dependency will be $k^{-3 + 2\nu_m}$.

By default, we use $N_0 = 5$ and $N_2 = 6$ for the asymptotic expansion, for which higher orders are negligible on scales $k \gtrsim 0.3 \, h \textrm{Mpc}^{-1}$ at $z = 0$ to about $1\%$ (see also figure~\ref{fig:ir-resum}).

\subsubsection{Combining the Methods}\label{sec:ir-combination}

The final expression for the IR resummation is a combination of the two methods presented above,
\begin{equation}
    \label{eq:IRResumPowerSpectrumTotal}
    P^{\ell}_{\textrm{IR}}(k;z) \Big\rvert_{\mathcal{N}} = P^{\ell}_{\textrm{IR,direct}}(k;z) \Big\rvert_{\mathcal{N}} + w(k) \bigg(P^{\ell}_{\textrm{IR,asympt}}(k;z) \Big\rvert_{\mathcal{N}} - P^{\ell}_{\textrm{IR,direct}}(k;z) \Big\rvert_{\mathcal{N}}\bigg)\,,
\end{equation}
with $P^{\ell}_{\textrm{IR}}(k; z)\rvert_{\mathcal{N}} = P^{\ell}(k; z)\rvert_{\mathcal{N}} - P^{\ell}(k; z)\rvert\rvert_{\mathcal{N}} \equiv P_{\textrm{IR}}(k; z)$ and where $w(k)$ is a window function that transitions between the two methods. By default, we choose $w(k)$ to be the same tapered window function we also use for the suppression of Gibbs phenomenon in the FFTLog coefficients, eq.~\eqref{eq:WindowFunction}. 

To apply the method of Section~\ref{sec:ap} for the analytical AP effect to this expression, an additional FFTLog is performed to obtain the required power series form of eq.~\eqref{eq:APEffectPowerSpectrumMultipolePowerLaw}, i.e., 
\begin{equation}
    P^{\ell}_{\textrm{IR}}\big(k; z\big) = k^{\nu} \sum_m \tilde{c}_m\big[P^{\ell}_{\textrm{IR}}\big](z) \, k^{\eta_m}\,.
    \label{eq:IRResumPowerSpectrumDecomposition}
\end{equation}

\begin{figure}[t]
    \centering
    \includegraphics[width=\textwidth]{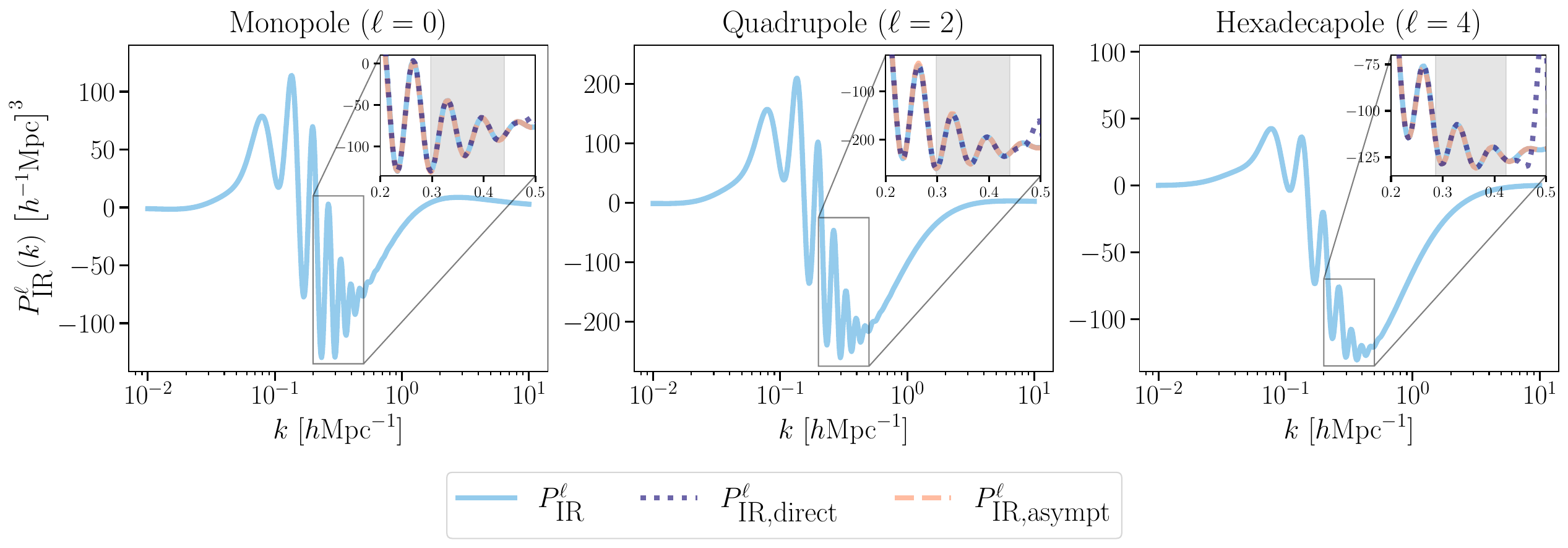}
    \caption{The IR resummed power spectrum $P^{\ell}_{\textrm{IR}}(k)$ with $\ell = 0,2,4$ in a $\Lambda$CDM cosmology at $z = 0$ with Planck best-fit values and $\Lambda_{\textrm{IR}} = 0.12 \, h/\textrm{Mpc}$. The grey zones in the zoomed-in regions show the transition between the direct and asymptotic methods.}
    \label{fig:ir-resum}
\end{figure}

The upper bound for the transition, $k_{\max}$, is determined automatically by testing for the divergence of the direct method on the logarithmic grid of the FFTLog at each multipole order individually. In particular, when viewing the expansion in eq.~\eqref{eq:IRResumPowerSpectrumDirect} as $P^{\ell}_{\textrm{IR,direct}}(k; z) = \sum_{n = 1}^{N} T^{\ell}_n(k; z)$, we determine convergence when for some $N^{\star} \leq N$
\begin{equation}
    \left|\frac{T^{\ell}_{N^{\star} - 1}(k; z)}{S^{\ell}_{N^{\star} - 1}(k; z)}\right| < \left|\frac{T^{\ell}_{N^{\star}}(k; z)}{S^{\ell}_{N^{\star}}(k; z)}\right| \quad \textrm{and} \quad \left|\frac{T^{\ell}_{N^{\star} - 1}(k; z)}{S^{\ell}_{N^{\star} - 1}(k; z)}\right| < 1\% \, ,
\end{equation}
where $S^{\ell}_{N^{\star}}(k; z) = \sum_{n = 1}^{N^{\star}} T^{\ell}_n(k; z)$ is the $N^{\star}$'s partial sum of $P^{\ell}_{\textrm{IR,direct}}(k; z)$. $k_{\max}$ is then chosen as the earliest instance on the logarithmic grid for which the conditions are not fulfilled, as well as where the direct method disagrees with the asymptotic method at more than $1\%$.

On the other hand, the lower bound for the transition, $k_{\min}$, is chosen statically, namely as the point on the logarithmic grid that lies $n^{\star}$ places below $k_{\max}$. I.e., $\log k_{\min} = \log k_{\max} - n^{\star} \delta \log k$, where $\delta \log k$ is the grid spacing of the logarithmic grid. By default, we use $n^{\star} = 10$. The main motivation for this is to have sufficiently many points in the FFTLog to ensure a smooth transition without potential discontinuities that could cause unwanted ringing.

We have checked that this procedure, together with the chosen orders for the perturbative expansions of $K_0(\vec{k}, \vec{q})$ in Sections~\ref{sec:ir-direct} and~\ref{sec:ir-asympt}, correctly chooses a region for the transition, in which the direct and asymptotic methods, except for zero-crossings, agree better than $\sim 1\%$ for Planck-like $\Lambda$CDM cosmologies (cf.~Section~\ref{sec:comparison}) at the present epoch, $z = 0$, as well as all the way up to redshift $z \sim 10$, where the IR resummation becomes negligible.

As an example, we show the IR resummed part of the power spectrum, $P^{\ell}_{\textrm{IR}}(k; z)$, for the case $z = 0$ in figure~\ref{fig:ir-resum} for the first three multpoles. In the zoomed-in regions, we further show the transition between the direct and asymptotic methods. As can be seen, the transition region slightly differs for each multipole, with higher multipole order transitioning earlier. This is a general observation that we make. Furthermore, slightly beyond each transition region, the direct method starts to diverge, which is to be expected for the criterion we chose for the upper bound of the transition region. The divergence, in this case, occurs roughly around $k \gtrsim 0.45 \, h\textrm{Mpc}^{-1}$. For larger redshifts, the linear growth rate $D_{+}(z)$ becomes smaller, causing the linear power spectrum $P^{(0)}(k; z)$, and therefore the filters $\Xi_0(q; z)$ and $\Xi_2(q; z)$ to reduce as well. The overall effect is the extension of the direct method to larger values of $k$, as the perturbative expansion of $K_0(\vec{k}, \vec{q}; z)$~\eqref{eq:IRResumKernelK} becomes more accurate.

\section{The \texttt{neckar} code}\label{sec:code}

The code accompanying this work, \texttt{neckar}, provides significant improvements in numerical precision and computational time of the one-loop power spectrum in the FFTLog framework with the various contributions that are common in modern cosmology of LSS, while at the same time being straightforward to use. To satisfy these needs, the fully serial code is written in C, with FFTW~\cite{Frigo:2005zln}\footnote{\href{https://fftw.org/}{https://fftw.org/}} and GSL~\cite{Galassi_2009}\footnote{\href{https://www.gnu.org/software/gsl/}{https://www.gnu.org/software/gsl/}} as the sole external dependencies, by adopting a modular and object-oriented philosophy. Equipped with a Cython wrapper to Python, \texttt{neckar} is naturally translated into Python, making access from within Python almost lossless, ideal for MCMC. Tutorials on the use of the code are readily available in the form of Jupyter notebooks in the repository.

\subsection{Products}

From the outside, \texttt{neckar} makes no distinction between the power spectrum and the correlation function: both are computed simultaneously. Apart from the stochastic power spectrum, which does not enter the correlation function, every other contribution to the power spectrum is expressed internally as an FFTLog,
\begin{equation}
    P^{\ell}(k; z)_j = \sum_{m} \mathcal{P}^{\ell}_m(z)_j \, k^{-2\nu_m} \, ,
    \label{eq:PowerSpectrumDecomposition}
\end{equation}
From the definition of the correlation function in eq.~\eqref{eq:IRResumCorrFunction}, each part of the power spectrum can therefore directly be translated into configurational space as
\begin{equation}
    \xi^{\ell}(q; z)_j = \sum_{m} \mathcal{X}^{\ell}_m(z)_j \, q^{-3 + 2\nu_m}\,,
    \label{eq:CorrelationFunctionDecomposition}
\end{equation}
with the coefficients of the correlation function given by
\begin{equation}
    \mathcal{X}^{\ell}_m(z)_j = \frac{4\pi i^{\ell}}{(2\pi)^3} \, I_3(\ell, 2 + 2\nu_m) \, \mathcal{P}^{\ell}_m(z)_j \,,
    \label{eq:CorrelationFunctionDecompositionCoefficients}
\end{equation}
where $I_3$ is the integral solution of eq.~\eqref{eq:AppIntJnPow}. Furthermore, thanks to the method of Section~\ref{sec:ap}, the AP effect is already imprinted on the power spectrum coefficients $(\mathcal{P}^{\ell}_m)_j$, such that no additional work needs to be done for the correlation function coefficients $(\mathcal{X}^{\ell}_m)_j$. In this sense, the correlation function can be viewed as a byproduct of the power spectrum that causes almost no additional overhead.

\subsection{Structure and Implementation}

\begin{table}[t]
    \renewcommand\arraystretch{1.4}
    \centering
    \begin{tabular}{c|ccccc|c}
        \hline\hline
        Object of the FFTLog & $\nu$ & $N$ & min & max & $x_{0r}$ & Pre-Computed \\ \hline
        $P^{(0)}(k)$ in $P_{\textrm{tree,SPT}}(k)$~\eqref{eq:Ptree} & -1.1 & 256 & -4 & 2 & 0.8 & \\  \hline
        $P^{(0)}(k)$ in $P_{\textrm{1-loop, SPT}}(k)$~\eqref{eq:P1loop} & \multirow{2}{*}{-1.6} & \multirow{2}{*}{256} & \multirow{2}{*}{-4} & \multirow{2}{*}{2} & \multirow{2}{*}{0.8} & $I$~\eqref{eq:P1loopMasterIntegral} \\ 
        and in $P_{\textrm{1-loop,UV}}(k)$~\eqref{eq:P1loopUV} & & & & & \\ \hline
        $P_{\textrm{IR}}(k)$~\eqref{eq:IRResumPowerSpectrumDecomposition} & 0.1 & 256 & -3 & 2 & 0.9 & \\ \hline
        $P^{(0)}(k)$ in $\Xi_0^{\infty}$~\eqref{eq:IRFilterXi0InfFFTLog} & \multirow{2}{*}{0} & \multirow{2}{*}{128} & \multirow{2}{*}{-4} & \multirow{2}{*}{2} & \multirow{2}{*}{0.8} & $I_2$~\eqref{eq:AppIntExpPow} \\ 
        and in $\Xi^{(2)}_0$~\eqref{eq:IRResumFilterXi0DerivFFTLog} & & & & & & $I_3$~\eqref{eq:AppIntJnPow} \\ \hline
        $P^{(0)}(k) \exp[-k^2/\Lambda_{\textrm{IR}}^2]$ in $\hat{\Xi}_0$~\eqref{eq:IRFilterHatXiFFTLog} & -0.2 & 128 & -4 & 2 & 0.8 & $I_3$~\eqref{eq:AppIntJnPow} \\ \hline
        $P^{(0)}(k) \exp[-k^2/\Lambda_{\textrm{IR}}^2]$ in $\hat{\Xi}_2$~\eqref{eq:IRFilterHatXiFFTLog} & -1.2 & 128 & -4 & 2 & 0.8 & $I_3$~\eqref{eq:AppIntJnPow} \\ \hline
        $\Xi_0^{n_0}(q) \Xi_2^{n_2}(q) \xi^{\ell}(q)_j$~\eqref{eq:IRResumMasterIntegralDirectFFTLog} & \multirow{2}{*}{-1.6} & \multirow{2}{*}{256} & \multirow{2}{*}{-1} & \multirow{2}{*}{4} & \multirow{2}{*}{0.9} & $I_3$~\eqref{eq:AppIntJnPow} \\
        and $\tilde{\Xi}_0^{n_0}(q) \tilde{\Xi}_2^{n_2}(q) \xi^{\ell}(q)_j$~\eqref{eq:IRResumMasterIntegralAsymptoteFFTLog} & & & & & & $I_4$~\eqref{eq:AppIntJnExpPow} \\
        \hline\hline
    \end{tabular}
    \caption{Default settings for the various FFTLogs that are performed in \texttt{neckar}, along with the integrals that are precomputed. Here, $\nu$ is the bias of the FFTLog, $N$ is the grid size, and $\min/\max$ correspond to $\log_{10} x_{\min/\max}$, with $x \equiv q$ [Mpc$/h$] or $x \equiv k$ [$h/$Mpc], the bounds of the grid on which the respective FFTLog is performed. For the mitigation of ringing, we use $x_0 = x_{0r} \, N/2$ and $x_1 = N/2$ for the window function of eq.~\eqref{eq:WindowFunction}.}
    \label{tab:fftlog}
\end{table}

For an efficient implementation of the one-loop integrals and the IR resummation, \texttt{neckar} distinguishes between three different phases: the \textit{configuration} phase, the \textit{setup} phase, and the \textit{evaluation} phase.

The configuration phase, apart from C-specific preparations such as memory allocation, is responsible for the pre-computation of various numerically expensive quantities. In fact, by having the FFTLog separate the parameter-dependence\footnote{By parameter-dependence, we mean the dependence of the power spectrum on any free parameter that would be explored in an MCMC analysis, i.e., the galaxy bias parameters or UV counter-term parameters, but also the cosmological parameters entering through $P^{(0)}(k; z)$, $f(z)$, $H(z)$, or $D_A(z)$.} of the numerous contributions to the power spectrum from their scale-dependence (cf.~eq.~\eqref{eq:PowerSpectrumDecomposition}), several computationally expensive integrals become pre-computable. Additionally to the integrals $I_3$~\eqref{eq:AppIntJnPow} needed for the transition of the single power spectrum contributions to configuration space in eq.~\eqref{eq:CorrelationFunctionDecompositionCoefficients}, we list the various other integrals that are pre-computed\footnote{Note that, in principle, $I_4(\alpha, \mu, \nu)$ enters with a cosmology-dependence in the $\alpha$ parameter, which is alleviated by the creation and subsequent interpolation of a large table of this integral for several $\alpha$ values. This is especially important in regard to the slow convergence of the ${}_1F_1$ hypergeometric function (cf. Appendix~\ref{app:HyperConfluent}).} in table~\ref{tab:fftlog} along with the default settings for the respective FFTLogs that are performed. Importantly, at this step, only the settings for the FFTLogs need to be specified; no knowledge of the free parameters is required.

On the other hand, the setup phase is responsible for combining the explicit parameter-dependence with the pre-computed integrals of the configuration phase to form the coefficients of the various power spectrum and correlation function contributions, $(\mathcal{P}^{\ell}_m)_j$ and $(\mathcal{X}^{\ell}_m)_j$, that appear in eq.~\eqref{eq:PowerSpectrumDecomposition} and eq.~\eqref{eq:CorrelationFunctionDecomposition}, respectively.

Among others, this step involves performing the FFTLogs of the input linear power spectrum $P^{(0)}(k)$ for the tree-level and one-loop order EFT contributions. In particular, $P^{(0)}(k)$ may be provided on an arbitrary grid, as it is internally mapped onto the FFTLog-specific logarithmic grid via cubic-spline interpolation and, if needed, power-law extrapolation to low/high $k$. The resulting FFTLog coefficients are then combined with the redshift-space  kernels of biased tracers and the pre-computed one-loop master integrals or the UV counter-term parameters to form the $\mu$-dependent $\mathcal{P}_{m}(\mu)_j$ coefficients of the EFT contributions as laid out in Sections~\ref{sec:spt} and~\ref{sec:uv}. Finally, by performing the multipole expansion~\eqref{eq:APEffectPowerSpectrumCoefMultipole} analytically, the $(\mathcal{P}^{\ell}_{m})_j$ coefficients are formed and further imprinted with the AP effect via the method of Section~\ref{sec:ap}, before additionally being translated into configuration space via eq.~\eqref{eq:CorrelationFunctionDecompositionCoefficients} with the pre-computed conversion integrals $I_3$.

For the IR resummation, additional FFTLogs of the same user-provided linear power spectrum are performed to form the IR filters $\Xi_0(q)$ and $\Xi_2(q)$ according to Section~\ref{sec:ir-filters}. After the computation of the coefficients $\mathcal{Q}^{\ell \ell'}_{n n_0 \alpha || \mathcal{N}-j}$ and $\tilde{\mathcal{Q}}^{\ell \ell'}_{n \tilde{n} \tilde{n}_0 \alpha || \mathcal{N}-j}$, the (filtered) correlation functions $\Xi^{n_0}_0(q) \Xi^{n_2}_2(q) \xi^{\ell}(q)_j$ and $\tilde{\Xi}^{\tilde{n}_0}_0(q) \tilde{\Xi}^{\tilde{n}_2}_2(q) \xi^{\ell}(q)_j$ are formed and then decomposed with FFTLogs, before combining all ingredients with the pre-computed integrals $I_3$ and $I_4$ according to Sections~\ref{sec:ir-direct} and~\ref{sec:ir-asympt} for the direct and asymptotic method, respectively. Lastly, the methods are combined according to Section~\ref{sec:ir-combination} in an FFTLog, before also being imprinted with the AP effect and translated into configuration space.

Finally, during the evaluation phase, coefficients of each contribution are combined with the explicit scale-dependence according to eqs.~\eqref{eq:PowerSpectrumDecomposition} and~\eqref{eq:CorrelationFunctionDecomposition} for the power spectrum and correlation function, respectively. The contributions are then combined to form the full one-loop predictions, $P^{\ell}(k; z) = \sum_{j} P^{\ell}(k; z)_j$ and $\xi^{\ell}(q; z) = \sum_{j} \xi^{\ell}(q; z)_j$. This step is the fastest, as it only involves linear operations.

\subsection{Performance}

\begin{figure}[t]
    \centering
    \includegraphics[width=0.94\textwidth]{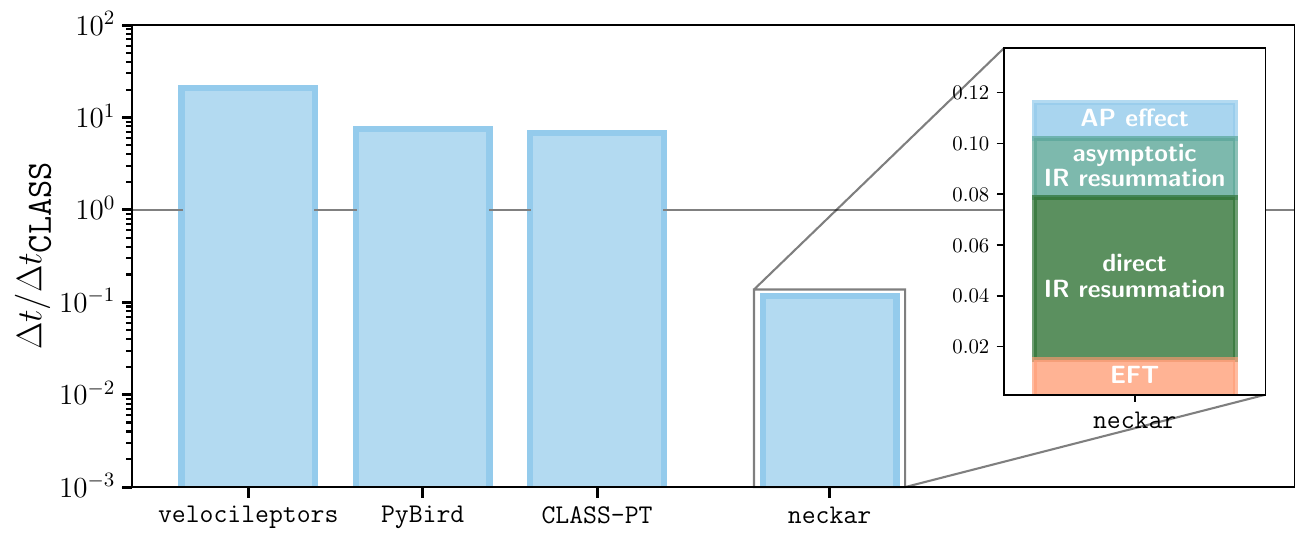}
    \caption{Execution times of popular codes with \texttt{CLASS} as the baseline: \texttt{velocileptors}, \texttt{PyBird}, and \texttt{CLASS-PT}, in comparison with \texttt{neckar}. The computation time of \texttt{neckar} only takes the \textit{setup} and \textit{evaluation} phases into account, and is further split into its most important contributions. We evaluate all codes on a grid of size $N_f = 128$ from $k_{f,\min} = 10^{-3} \, h \textrm{Mpc}^{-1}$ to $k_{f,\max} = 0.2 \, h \textrm{Mpc}^{-1}$. This figure makes it clear that the bottleneck is now the CLASS runs.}
    \label{fig:execution-times}
\end{figure}

We compare the performance of \texttt{neckar} in terms of execution time with various popular codes in figure~\ref{fig:execution-times} with \texttt{CLASS} as the baseline, which provided the linear power spectrum on a grid of size $N_i = 512$ from $k_{i,\min} = 10^{-4} \, h \textrm{Mpc}^{-1}$ to $k_{i,\max} = 10 \, h \textrm{Mpc}^{-1}$ for the Planck-like $\Lambda$CDM cosmology at $z = 0$ described in Section~\ref{sec:comparison}. Each code was instructed to compute the full one-loop results with IR resummation and AP effect for all three major multipoles of the power spectrum, monopole, quadrupole, and hexadecapole, evaluated on a grid of size $N_f = 128$ from $k_{f,\min} = 10^{-3} \, h \textrm{Mpc}^{-1}$ to $k_{f,\max} = 0.2 \, h \textrm{Mpc}^{-1}$.

From our test, it is apparent that \texttt{neckar} is about $\sim 60$ times faster than \texttt{PyBird} and \texttt{CLASS-PT}, and more than 100 times faster than \texttt{velocileptors}. Furthermore, since \texttt{neckar} is also considerably faster than \texttt{CLASS}, cosmology-dependent MCMC-analyses of the full-shape power spectrum with \texttt{neckar} will be bottlenecked by \texttt{CLASS}. 

Considering the breakdown of the single contributions to the total execution time of \texttt{neckar}, the IR resummation is by far the biggest contributor, requiring roughly $\sim 75\%$ of the total execution time. On the other hand, both the EFT calculations and the inclusion of the AP effect require only about $\sim 12.5\%$ of the total execution time each. The $\sim 12.5\%$ for the EFT calculations includes the tree-level and one-loop order calculations for the full $\mu$-dependence of the kernels of Appendix~\ref{app:Model}, as well as the extraction of the first three multipoles and the subsequent evaluation. Although not shown, of the EFT calculation time, roughly one third is spent in the evaluation phase, and only two thirds in the actual setup phase, with the latter translating to $\sim 1 \, \textrm{ms}$. The evaluation of the IR resummation takes roughly a quarter of that for the EFT calculations, as only a single FFTLog needs to be evaluated.

\section{Comparisons with \texttt{PyBird}}\label{sec:comparison}

In this section we make qualitative comparisons with \texttt{PyBird}, as it is not only a popular code in the literature, but it also has been the qualitative reference in the early phases of development of our code, especially in regard to the IR resummation. In fact, direct qualitative comparisons between \texttt{neckar} and \texttt{velocileptors} or \texttt{CLASS-PT} are not possible due to differing approaches to the IR resummation (e.g., the wiggle/non-wiggle method~\cite{Blas:2016sfa}). Still, we confirmed that in absence of IR resummation, and where applicable,\footnote{\texttt{CLASS-PT} does not allow the AP effect to enter independently of the IR resummation, while \texttt{velocileptors} (LPT) always has the IR resummation enabled, which one can only ``turn'' off by pushing $\Lambda_{\textrm{IR}} \sim 0.001 \, \textrm{Mpc}/h$.} the other codes agree with our code to well below 1\% in precision.

For the comparisons, we use version 0.3.1 of \texttt{PyBird}. We assume no galaxy bias and no stochastic/counter terms,\footnote{We tested that no noticeable differences arise from the inclusion of either terms.} and use a Planck-like $\Lambda$CDM cosmology at $z = 0$ with~\cite{Planck:2018vyg}
\[
    (\ln(10^{10} A_s), n_s, h, \omega_b, \omega_{cdm}) = (3.045, 0.9661, 0.6732, 0.02238, 0.1201)
\]
for the linear power spectrum computed with \texttt{CLASS}, as well as $f(z) = \Omega_m(z)^{0.545}$ for the growth rate. To ``cheat'' in an AP effect, we use a non-zero redshift, $z_{\textrm{AP}} = 0.31$, and adopt $E^2(z_{\textrm{AP}}) = H^2(z_{\textrm{AP}}) / H^2_0 = \Omega^{(0)}_m (1 + z_{\textrm{AP}})^3 + 1 - \Omega^{(0)}_m$ for the expansion rate and $L_A(z_{\textrm{AP}}) = H_0 D_A(z_{\textrm{AP}}) = (1 + z_{\textrm{AP}})^{-1} \int^{z_{AP}}_0 dz\, E^{-1}(z)$ for the (dimensionless) angular diameter distance, with $(\Omega^{(0)}_m)' = 0.2$ for the fiducial values $E'(z_{\textrm{AP}})$ and $L_A'(z_{\textrm{AP}})$.

\subsection{Without IR Resummation}

\begin{figure}[t]
    \centering
    \includegraphics[width=\textwidth]{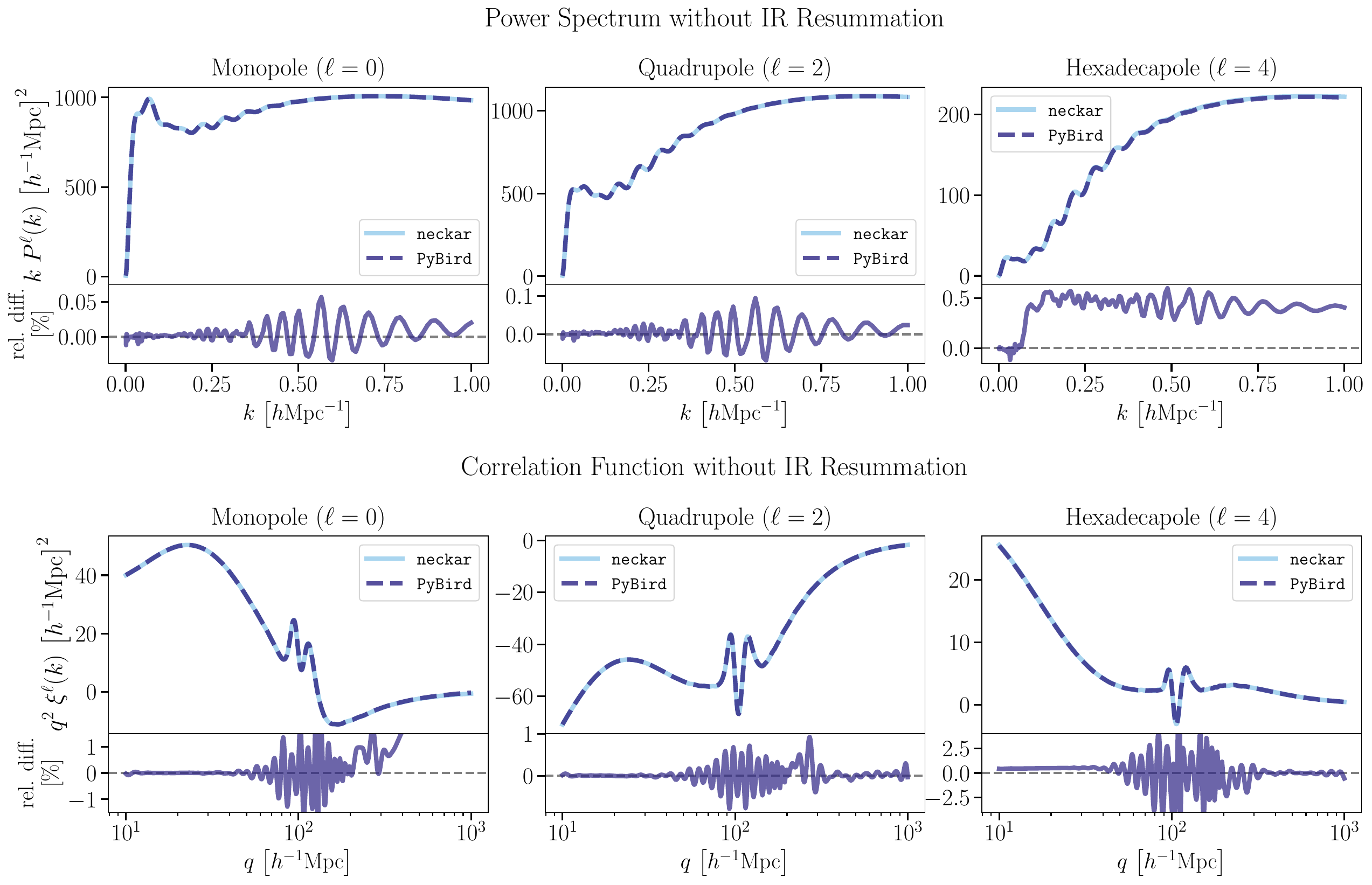}
    \caption{Comparison of $P^\ell(k)$ (top) and $\xi^{\ell}(q)$ (bottom) of \texttt{neckar} and \texttt{PyBird} without IR resummation for our fiducial cosmology (see text). Each plot also shows the relative difference between the results of the two codes. For the correlation function residual plots we choose the scale
    to highlight the differences around the BAO peak.}
    \label{fig:pkfft-vs-pybird-withoutIR}
\end{figure}

In absence of the IR resummation, we first show figure~\ref{fig:pkfft-vs-pybird-withoutIR} as a qualitative comparison not only of the entire FFTLog machinery for the one-loop integrals, but also of the AP effect, which, to remind the reader, is fully semi-analytical in our approach.

From the top-row, where we show the power spectrum multipoles, the agreement can be observed to be well below $1\%$, with the monopole and quadrupole even agreeing to well below $0.1\%$. Slightly higher deviations ($\sim0.5\%$) are apparent in the hexadecapole. These deviations are an artifact of the premature truncation of the multipole expansion of eq.~\eqref{eq:APEffectPowerSpectrumDecomposition} at $\ell = 4$ in \texttt{PyBird}, when, in fact, also multipoles of orders $\ell = 6$ and $\ell = 8$ enter the power spectrum with the kernels of Appendix~\ref{app:RSDKernels}. Still, at deviations well below $1\%$, this truncation is more than justified.

The bottom-row instead shows the correlation function multipoles. While mostly agreeing to better than $1\%$, deviations in the monopole on large scales, $q \to 10^3 \, h^{-1} \textrm{Mpc}$ can be explained by our use of an additional FFTLog of $P^{(0)}(k)$ for the tree-level power spectrum. By choosing a different bias (cf.~table~\ref{tab:fftlog}) than what is used for the one-loop power spectrum, the small $k$ behaviour of $P^{(0)}(k)$ is better resolved, and hence the large $q$ behaviour of the correlation function at tree-level. Additionally, $1\%$-level deviations around the BAO peak, apart from the zero-crossings in the monopole and hexadecapole, largely originate from ringing defects and the different filters used in the codes to mitigate them.

\subsection{With IR Resummation}

\begin{figure}[t]
    \centering
    \includegraphics[width=\textwidth]{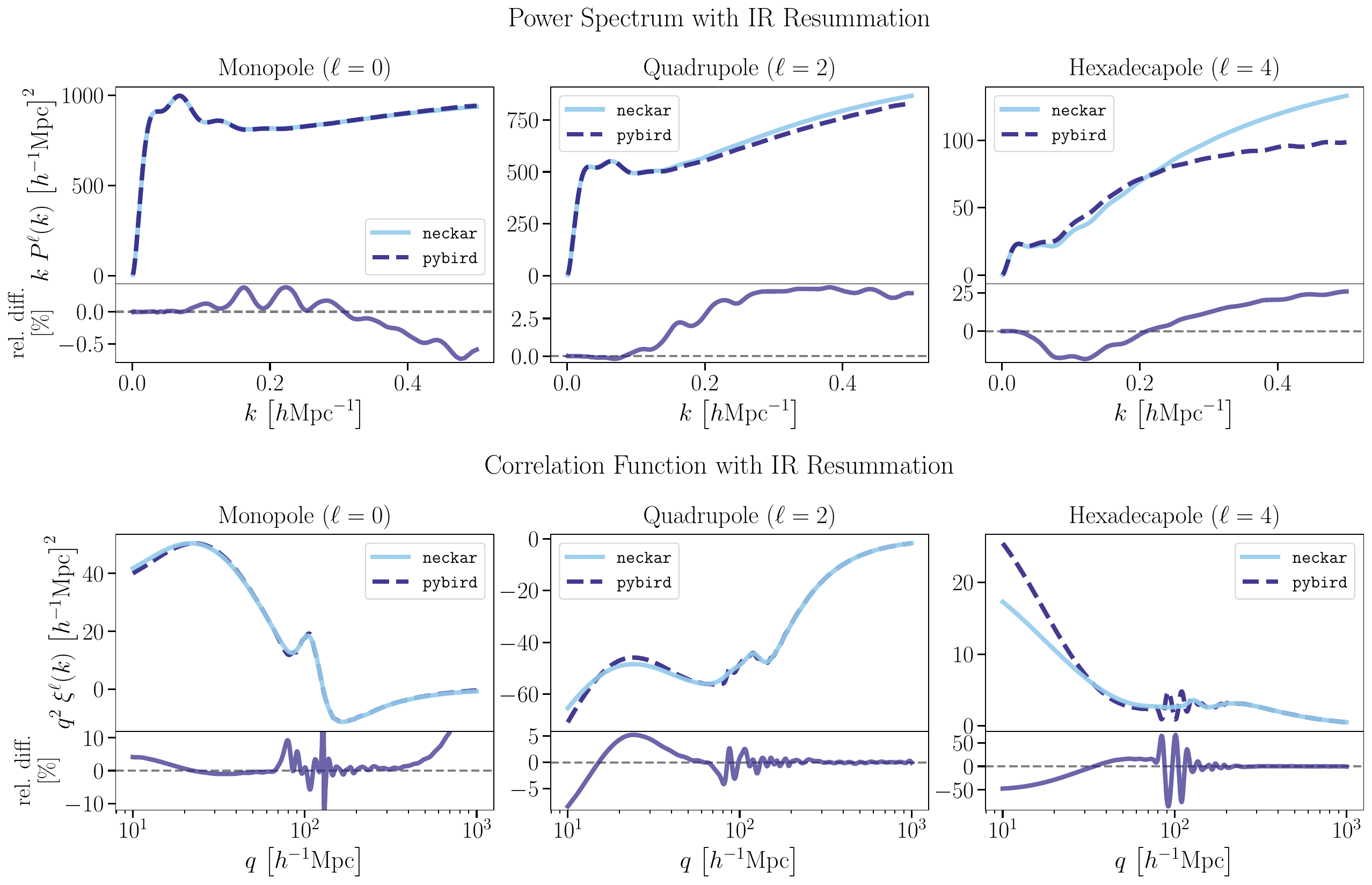}
    \caption{ 
    Same as figure~\ref{fig:pkfft-vs-pybird-withoutIR}
    but now including the IR resummation.
    }
    \label{fig:pkfft-vs-pybird-withIR}
\end{figure}

When including the IR resummation, the results of the power spectrum multipoles start to deviate substantially for growing multipole order, as we show in the top-row of figure~\ref{fig:pkfft-vs-pybird-withIR}. While the monopole still agrees to below $1\%$, the quadrupole starts to show disagreements at $1\%$-order already on mildly non-linear scales $k \gtrsim 0.15 \, h \textrm{Mpc}^{-1}$, and the hexadecapole even disagrees at $10\%$-level on scales of $k \gtrsim 0.1 \, h \textrm{Mpc}^{-1}$.

As was shown in figure~\ref{fig:ir-resum}, the direct method still agrees with the asymptotic method well beyond $k \gtrsim 0.4 \, h \textrm{Mpc}^{-1}$. Furthermore, since the asymptotic method is not used on scales $k < 0.25 \, h \textrm{Mpc}^{-1}$, the discrepancy must lie with \texttt{PyBird}'s implementation of the IR resummation. In fact, there are a total of three inconsistencies in the implementation of the IR resummation in \texttt{PyBird} that cause these deviations.

\begin{enumerate}
    \item \textit{Premature truncation of the $K_0(\vec{k}, \vec{q})$ expansion:} The least affecting inconsistency is the insufficient choice of $N_{0a} = 14$, $N_{0b} = 5$, and $N_2 = 1$ in \texttt{PyBird}. The value of $N_2$, in particular, is too low, which manifests itself as $\sim 0.5\%$ wiggles in the residual plots of both the monopole and quadrupole. We use instead $N_{0a} = 5$, $N_{0b} = 6$, and $N_2 = 7$.

    \item \textit{Premature truncation of expansion in $j_\alpha$:} In eq.~\eqref{eq:IRResumPowerSpectrumDirect}, 
    Bessel functions enter, in principle, with orders up to $\alpha_{\max} = \ell + 2(N_{0a} + N_{0b})$ for a given multipole order in the IR resummation. In \texttt{PyBird}, however, the choice is to truncate at $\alpha_{\max} = 4$. While the monopole and quadrupole are less affected by this choice, we find this truncation to be responsible for roughly half of the discrepancies beyond $k \gtrsim 0.2 \, h/\textrm{Mpc}$ in the hexadecapole. Furthermore, we find that choosing $\alpha_{\max} = 6$ is already sufficient to correct for the deviations. Our choice is to use $\alpha_{\max} = 10$, as we additionally include the multipoles $\ell = 6$ and $\ell = 8$ in the IR resummation.

    \item \textit{Missing Bessel functions:} Lastly, and most prevalently for the quadrupole and hexadecapole, in the process of extracting the coefficients $\mathcal{Q}^{\ell \ell'}_{n n_0 \alpha || \mathcal{N}-j}$ for eq.~\eqref{eq:IRResumPowerSpectrumDirect}, due to Bessel functions of orders $\alpha \ge 6$ being neglected in \texttt{PyBird}, contributions from implicit $j_{\alpha}$ of orders $\alpha \leq 4$ are missing. To elaborate, we emphasize that the $j_{\alpha}$ in eq.~\eqref{eq:IRResumPowerSpectrumDirect} result from derivatives of Bessel functions (cf.~eq.~\eqref{eq:AppIRResumBesselDerivative} in Appendix~\ref{app:ir-direct}). Since the coefficients in \texttt{PyBird} are precomputed in Mathematica, derivatives of Bessel functions there usually stem from the following relation\footnote{One can verify that this expression is equivalent to eq.~\eqref{eq:AppIRResumBesselFirstDerivative} by using eq.~\eqref{eq:BesselRecursion}.}
    \begin{equation}
        \frac{d}{dx} j_{\ell}(x) = -\frac{j_{\ell}(x)}{2x} + \frac{j_{\ell-1}(x) + j_{\ell+1}(x)}{2} \, ,
    \end{equation}
    producing Bessel functions of both even and odd orders in combination with their argument, $x$. Since only even Bessel functions without the explicit $x$ dependency must remain in eq.~\eqref{eq:IRResumPowerSpectrumDirect}, a repeated application of (see eq. 10.1.19 in~\cite{AbramowitzStegun})
    \begin{equation}
        j_{\ell - 1}(x) + j_{\ell + 1}(x) = \frac{2\ell + 1}{x} j_{\ell}(x) \, ,
        \label{eq:BesselRecursion}
    \end{equation}
    appropriately yields the desired result. However, neglecting all $j_\alpha$ of orders $\alpha \ge 6$, also contributions to $j_\alpha$ for $\alpha \leq 4$ get neglected, as they are hidden in this relation recursively. We circumvent this issue by using eq.~\eqref{eq:AppIRResumBesselFirstDerivative} for the derivatives, which produces Bessel functions of the same parity without an explicit $x$ dependency.
\end{enumerate}

These inconsistencies are overshadowed by ringing defects in the correlation function, which we show in the bottom-row of figure~\ref{fig:pkfft-vs-pybird-withIR}. In fact, since the IR resummation is only taken into account for the BAO peak in \texttt{PyBird}, ringing defects from the ad-hoc truncation of the resummation, which already appear on the monopole,\footnote{For higher redshifts, where non-linear effects are delayed to larger values of $k$, the direct IR resummation extends further. In this case, also \texttt{PyBird} is able to produce a smooth BAO peak in the monopole, however, generally not in the quadrupole and hexadecapole.} obscure the inconsistencies. In \texttt{neckar}, thanks to the asymptotic method's smooth extension to non-linear scales, such ringing defects are strongly suppressed. Moreover, corrections to the 
$\xi(q)$ on scales of $q < 100 \, h^{-1} \textrm{Mpc}$ are a direct consequence of the IR resummation's extension to non-linear scales.

\section{Discussion}\label{sec:conclusions}

In this work we introduced a new code, \texttt{neckar}, for the redshift-space one-loop power spectrum of biased tracers in the EFTofLSS. Adopting the FFTLog methodology of~\cite{Simonovic:2017mhp}, we additionally developed an efficient, FFTLog-compatible inclusion of the AP effect for the power spectrum multipoles, making \texttt{neckar} fully semi-analytical. With this, and thanks to an efficient implementation in the C-programming language, \texttt{neckar} is considerably faster than previous similar codes by a factor of more than $\sim 60$. At the same time, we improved the implementation of the IR resummation in \texttt{PyBird}~\cite{DAmico:2020kxu} on linear and mildly non-linear scales $k \lesssim 0.4 \, h \textrm{Mpc}^{-1}$ in both efficiency and precision, which we dub the ``direct'' method, and we provided a smooth extension into the non-linear regime, which we call the ``asymptotic'' method. We also showed the benefits of this extension for the multipoles of the correlation function, providing a smooth and noise reduced BAO peak.\footnote{Note that~\cite{Senatore:2017pbn} provided a method for the IR resummation directly for the correlation function using the FFTLog. Similarly to our extension to non-linear scales,
they also obtain a functional form that includes a confluent hypergeometric function, although of the second kind, $U(a; b; z)$.}

An extension of the code to estimate the tree-level bispectrum, in principle, is straightforward, as the relevant kernels are already present (cf.~eq.~\eqref{eq:AppRSDKernelsZ2}). However, to apply the AP effect to the bispectrum multipoles semi-analytically via the FFTLog, even when focusing on just the monopole, the method of section~\ref{sec:ap} would need to be extended. On the other hand, the method of section~\ref{sec:ap} could directly be used for an extension to $P(k)$ at two-loop order, for which the IR resummation is also applicable. By exploring the method of~\cite{Simonovic:2017mhp} for the two-loop order master integrals, it may be possible to retain a significant amount of efficiency while gaining access to smaller, more non-linear scales.

At the same time, we plan to apply the code to analyse real data provided by current surveys. We expect \texttt{neckar} to be particularly useful the more model-independent the analysis is, due to the large number of parameters that need to be introduced. These tasks will be explored in future works.

\medskip

\section*{Appendix}

\appendix

\section{Theoretical Model}\label{app:Model}

\subsection{Cosmology Kernels}\label{app:CosmKernels}

From the following base of functions (see~\cite{Bernardeau:2001qr} for definitions),
\begin{subequations}
    \begin{align}
        \alpha\big(\vec{k}_1, \vec{k}_2\big) & = 1 + \frac{\vec{k}_1 \cdot \vec{k}_2}{k_1^2}\\
        \beta\big(\vec{k}_1, \vec{k}_2\big) & = \frac{\vec{k}_1 \cdot \vec{k}_2}{2 k_1 k_2} \bigg(\frac{k_1}{k_2} + \frac{k_2}{k_1}\bigg) + \bigg(\frac{\vec{k}_1 \cdot \vec{k}_2}{k_1 k_2}\bigg)^2 \\
        \gamma\big(\vec{k}_1, \vec{k}_2\big) & = 1 - \bigg(\frac{\vec{k}_1 \cdot \vec{k}_2}{k_1 k_2}\bigg)^2,
    \end{align}
\end{subequations}
and together with the anti-symmetric part of $\alpha(\vec{k}_1, \vec{k}_2)$,
\begin{equation}
    \alpha^{(a)} = \frac{\vec{k}_1 \cdot \vec{k}_2}{2 k_1 k_2} \bigg(\frac{k_2}{k_1} - \frac{k_1}{k_2}\bigg),
\end{equation}
the matter and velocity kernels considered here are constructed following the bootstrap approach of~\cite{DAmico:2021rdb}, which uses symmetries to parametrise the kernels for a class of cosmologies following those symmetries. By using the slightly different parametrisation of~\cite{Amendola:2023awr}, the kernels up to third order in perturbation theory read
\begin{subequations}
    \begin{equation}
        X^{(s)}_1\big(\vec{k}_1; z\big) = 1\,,
    \end{equation}
    \begin{equation}
        X^{(s)}_2\big(\vec{k}_1, \vec{k}_2; z\big) = \beta\big(\vec{k}_1, \vec{k}_2\big) + \frac{y^{(2)}_{\gamma}(z)}{2} \gamma\big(\vec{k}_1, \vec{k}_2\big)\,,
    \label{eq:AppModelKernelSecondOrder}
    \end{equation}
    \begin{align}
        X^{(s)}_3\big(\vec{k}_1, \vec{k}_2, \vec{k}_3; z\big) &= \frac{1}{3} \bigg[\beta\big(\vec{k}_{12}, \vec{k}_3\big) \beta\big(\vec{k}_1, \vec{k}_2\big) + \cyc\bigg] \nonumber \\
        &+ \frac{1}{6} \bigg(\frac{1}{2} y^{(3)}_{\gamma a}(z) + y^{(3)}_{\gamma b}(z)\bigg) \bigg[\gamma\big(\vec{k}_{12}, \vec{k}_3\big) \gamma\big(\vec{k}_1, \vec{k}_2\big) + \cyc\bigg] \nonumber \\
        &+ \frac{1}{6} \bigg(\frac{1}{2} y^{(3)}_{\gamma a}(z) - y^{(3)}_{\gamma b}(z)\bigg) \bigg[\alpha^{(a)}\big(\vec{k}_{12}, \vec{k}_3\big) \gamma\big(\vec{k}_1, \vec{k}_2\big) + \cyc\bigg] \label{eq:AppModelKernelThirdOrderPre} \\
        &- \frac{1}{6} \bigg(\frac{1}{2} y^{(3)}_{\gamma a}(z) - y^{(3)}_{\gamma b}(z) - 2h(z)\bigg) \bigg[\beta\big(\vec{k}_{12}, \vec{k}_3\big) \gamma\big(\vec{k}_1, \vec{k}_2\big) + \cyc\bigg] \nonumber \\
        &+ \frac{1}{3} \bigg(\frac{1}{2} y^{(3)}_{\gamma a}(z) - y^{(3)}_{\gamma b}(z) + y^{(2)}_{\gamma}(z) - h(z)\bigg) \bigg[\gamma\big(\vec{k}_{12}, \vec{k}_3\big) \beta\big(\vec{k}_1, \vec{k}_2\big) + \cyc\bigg], \nonumber
    \end{align}
\end{subequations}
where $\vec{k}_{1 \dots n} = \vec{k}_1 + \dots + \vec{k}_n$, and where $X^{(s)}_i \in \{F^{(s)}_i, G^{(s)}_i\}$ with $F^{(s)}_i$ and $G^{(s)}_i$ being the matter and velocity kernels, respectively, while $y^{(i)}_{\gamma x} \in \{a^{(i)}_{\gamma x}, d^{(i)}_{\gamma x}\}$ with $a^{(i)}_{\gamma x}$ and $d^{(i)}_{\gamma x}$ being the bootstrap parameters of the matter and velocity kernels, respectively, and
\begin{equation}
    h(z) = \int_{z}^{\infty} \frac{dz'}{1 + z'} f_{+}(z') \bigg[\frac{D_{+}(z')}{D_{+}(z)}\bigg]^2 d^{(2)}_{\gamma}(z').
    \label{eq:AppModelBootstrapHFunction}
\end{equation}

Note that eq.~\eqref{eq:AppModelKernelThirdOrderPre} enters the one-loop power spectrum only as $X^{(s)}_3(\vec{k}, \vec{q}, -\vec{q})$ (cf. eq.~\eqref{eq:P1loop}), simplifying the expression considerably in the process,\footnote{This expression can be derived from eq.~\eqref{eq:AppModelKernelThirdOrderPre} by using that
\[
    \beta\big(\vec{k}_1, \vec{k}_2\big) + \gamma\big(\vec{k}_1, \vec{k}_2\big) = \alpha^{(s)}\big(\vec{k}_1, \vec{k}_2\big)
\]
is the symmetric part of $\alpha(\vec{k}_1, \vec{k}_2)$, and
\[
    \alpha\big(\vec{q}, \vec{k} - \vec{q}\big) + \alpha\big(-\vec{q}, \vec{k} + \vec{q}\big) = 0
\]
\[
    \gamma\big(\vec{k} - \vec{q}, \vec{q}\big) \beta\big(\vec{k}, -\vec{q}\big) + \gamma\big(\vec{k} + \vec{q}, -\vec{q}\big) \beta\big(\vec{k}, \vec{q}\big) = 0.
\]}
\begin{equation}
    \begin{aligned}
        X^{(s)}_3\big(\vec{k}, \vec{q}, -\vec{q}; z\big) &= \frac{1}{3} \bigg[\beta\big(\vec{k} - \vec{q}, \vec{q}\big) \beta\big(\vec{k}, -\vec{q}\big) + \beta\big(\vec{k} + \vec{q}, -\vec{q}\big) \beta\big(\vec{k}, \vec{q}\big)\bigg] \\
        &+ \frac{y^{(3)}_{\gamma a}(z)}{6} \bigg[\gamma\big(\vec{k} - \vec{q}, \vec{q}\big) \gamma\big(\vec{k}, -\vec{q}\big) + \gamma\big(\vec{k} + \vec{q}, -\vec{q}\big) \gamma\big(\vec{k}, \vec{q}\big)\bigg] \\
        &+ \frac{h(z)}{3} \bigg[\beta\big(\vec{k} - \vec{q}, \vec{q}\big) \gamma\big(\vec{k}, -\vec{q}\big) + \beta\big(\vec{k} + \vec{q}, -\vec{q}\big) \gamma\big(\vec{k}, \vec{q}\big)\bigg],
    \end{aligned}
    \label{eq:AppModelKernelThirdOrder}
\end{equation}
while also eliminating the dependence on $y^{(3)}_{\gamma b}$. The usual approximation for these kernels is their well known solution in an Einstein-de Sitter (EdS) universe \cite{Bernardeau:2001qr, Goroff:1986ep}, for which the bootstrap parameters become constants, $a^{(2)}_{\gamma, \textrm{EdS}} = 10 / 7$, $d^{(2)}_{\gamma, \textrm{EdS}} = 6 / 7$, $a^{(3)}_{\gamma a, \textrm{EdS}} = 2 / 3$, $a^{(3)}_{\gamma b, \textrm{EdS}} = 5 / 9$, $d^{(3)}_{\gamma a, \textrm{EdS}} = 2 / 7$, $d^{(3)}_{\gamma b, \textrm{EdS}} = 5 / 21$, and $h_{\textrm{EdS}} = 3 / 7$ \cite{Amendola:2023awr}.

\subsection{Kernels of Biased Tracers}\label{app:BiasKernels}

\subsubsection{Galileons}

The default basis for biased tracers in \texttt{neckar} is in terms of Galilean invariant operators, dubbed the \textit{Galileons}, that have been introduced in~\cite{Chan:2012jj, Eggemeier:2018qae} with the parametrisation of~\cite{delaBella:2017qjy},
\begin{align}
    \delta_g\big(\vec{k}; z\big) &= b_1(z) \delta\big(\vec{k}; z\big) + \frac{b_2(z)}{2} \delta^2\big(\vec{k}; z\big) + \frac{b_3(z)}{6} \delta^3\big(\vec{k}; z\big) + b_{\mathcal{G}_2}(z) \mathcal{G}_2[\Phi_g]\big(\vec{k}; z\big) \label{eq:AppModelGalaxyBias} \\
    &+ b_{\mathcal{G}_2 \delta}(z) \delta\big(\vec{k}; z\big) \mathcal{G}_2[\Phi_g]\big(\vec{k}; z\big) + b_{\mathcal{G}_3}(z) \mathcal{G}_3[\Phi_g]\big(\vec{k}; z\big) + b_{\Gamma_3}(z) \Gamma_3[\Phi_g, \Phi_v]\big(\vec{k}; z\big) + ... \nonumber
\end{align}
where the ellipsis indicates higher order operators that do not enter at one-loop order, and where $\Phi_g = \nabla^{-2} \delta$ and $\Phi_v = \nabla^{-2} \theta$ are the potentials of the density contrast and the (scalar) velocity field $\theta = \nabla \cdot \vec{v}$, respectively, as well as
\begin{subequations}
    \begin{align}
        \mathcal{G}_2[\Phi] &= \big(\nabla_{ij} \Phi\big)^2 - \big(\nabla^2 \Phi\big)^2 \\
        \mathcal{G}_3[\Phi] &= \big(\nabla^2 \Phi\big)^3 + 2 \big(\nabla_{ij} \Phi\big) \big(\nabla_{jk} \Phi\big) \big(\nabla_{ki} \Phi\big) - 3 \big(\nabla_{ij} \Phi\big)^2 \big(\nabla^2 \Phi\big) \\
        \Gamma_3[\Phi_1, \Phi_2] &= \mathcal{G}_2[\Phi_1] - \mathcal{G}_2[\Phi_2]
    \end{align}
\end{subequations}
are the Galilean invariant tidal fields, where $\nabla_{ij} = \partial_i \partial_j$. The kernels of the biased tracers then follow as,
\begin{subequations}
    \begin{equation}
        K^{(s)}_1\big(\vec{k}_1; z\big) = b_1(z)
    \end{equation}
    \begin{equation}
        K^{(s)}_2\big(\vec{k}_1, \vec{k}_2; z\big) = b_1(z) F^{(s)}_2\big(\vec{k}_1, \vec{k}_2; z\big) - b_{\mathcal{G}_2}(z) \gamma\big(\vec{k}_1, \vec{k}_2\big) + \frac{b_2(z)}{2}
    \label{eq:AppModelBiasedKernelSecondOrder}
    \end{equation}
    \begin{equation}
        \begin{aligned}
            K^{(s)}_3\big(\vec{k}_1, \vec{k}_2, \vec{k}_3; z\big) &= b_1(z) F^{(s)}_3\big(\vec{k}_1, \vec{k}_2, \vec{k}_3; z\big) + \frac{b_2(z)}{3} \bigg[F^{(s)}_2\big(\vec{k}_1, \vec{k}_2; z\big) + \cyc \bigg] + \frac{b_3(z)}{6} \\
            &- \frac{2 b_{\mathcal{G}_2}(z)}{3} \bigg[\gamma\big(\vec{k}_{12}, \vec{k}_3\big) F^{(s)}_2\big(\vec{k}_1, \vec{k}_2; z\big) + \cyc \bigg] \\
            &- \frac{b_{\mathcal{G}_2 \delta}(z)}{3} \bigg[\gamma\big(\vec{k}_1, \vec{k}_2\big) + \cyc\bigg] + b_{\mathcal{G}_3}(z) L\big(\vec{k}_1, \vec{k}_2, \vec{k}_3\big) \\
            &- \frac{2 b_{\Gamma_3}(z)}{3} \bigg[\gamma\big(\vec{k}_{12}, \vec{k}_3\big) \bigg(F^{(s)}_2\big(\vec{k}_1, \vec{k}_2; z\big) - G^{(s)}_2\big(\vec{k}_1, \vec{k}_2; z\big)\bigg) + \cyc \bigg],
        \end{aligned}
        \label{eq:AppModelBiasedKernelThirdOrderPre}
    \end{equation}
\end{subequations}
where
\begin{equation}
    L\big(\vec{k}_1, \vec{k}_2, \vec{k}_3\big) = \bigg[\gamma\big(\vec{k}_1, \vec{k}_2\big) + \cyc \bigg] + 2 \Bigg[\bigg(\frac{\vec{k}_1 \cdot \vec{k}_2}{k_1 k_2}\bigg) \bigg(\frac{\vec{k}_1 \cdot \vec{k}_3}{k_1 k_3}\bigg) \bigg(\frac{\vec{k}_2 \cdot \vec{k}_3}{k_2 k_3}\bigg) - 1\Bigg].
\end{equation}

When considering the power spectrum at one-loop order, the last two terms in the first line of eq.~\eqref{eq:AppModelBiasedKernelThirdOrderPre}, as well as the first term in the third line, enter the renormalisation of $b_1$ at tree-level~\cite{Eggemeier:2018qae}. Furthermore, since the third order kernel enters as $K^{(s)}_3(\vec{k}, \vec{q}, -\vec{q}; z)$, not only is $L(\vec{k}, \vec{q}, -\vec{q}) = 0$, thus eliminating the $b_{\mathcal{G}_3}$ dependence, but the remainder also considerably simplifies,
\begin{equation}
    \begin{aligned}
        K^{(s)}_3\big(\vec{k}, \vec{q}, -\vec{q}; z\big) &= b_1(z) F^{(s)}_3\big(\vec{k}, \vec{q}, -\vec{q}; z\big) \\
        &- \frac{1}{3} b_{\times}(z) \bigg[\gamma\big(\vec{k} - \vec{q}, \vec{q}\big) \gamma\big(\vec{k}, -\vec{q}\big) + \gamma\big(\vec{k} + \vec{q}, -\vec{q}\big) \gamma\big(\vec{k}, \vec{q}\big)\bigg]
    \end{aligned}
    \label{eq:AppModelBiasedKernelThirdOrder}
\end{equation}
with
\begin{equation}
    b_{\times}(z) = b_{\mathcal{G}_2}(z) a^{(2)}_{\gamma}(z) + b_{\Gamma_3}(z) \Big(a^{(2)}_{\gamma}(z) - d^{(2)}_{\gamma}(z)\Big).
\end{equation}

\subsubsection{Basis of Descendants}

Another commonly used basis in the literature is the \textit{basis of descendants}~\cite{Angulo:2015eqa, Fujita:2016dne, Senatore:2014eva}, which we provide as an additional option in the code. We give here the relation to the galileons in the bootstrap cosmology as defined in Section~\ref{app:CosmKernels}. With the notation of~\cite{Perko:2016puo}, the kernels of biased tracers in the basis of descendants read\footnote
{
In particular, the following substitutions are made as compared to~\cite{Angulo:2015eqa, Fujita:2016dne}: $b^{\textrm{bod}}_1 = \tilde{c}_{\delta, 1}$, $b^{\textrm{bod}}_2 = \tilde{c}_{\delta, 2(2)}$, $\tilde{b}^{\textrm{bod}}_3 = \tilde{c}_{\delta, 3}$, $b^{\textrm{bod}}_4 = \tilde{c}_{\delta^2, 1(2)}$, and $\tilde{b}^{\textrm{bod}}_5 = \tilde{c}_{s^2, 2}$ (see also eq. (B.7) in~\cite{Perko:2016puo}).
}
\begin{subequations}
    \begin{equation}
        K^{(s)}_1\big(\vec{k}_1; z\big) = b^{\textrm{bod}}_1(z)
    \end{equation}
    \begin{equation}
        \begin{aligned}
            K^{(s)}_2\big(\vec{k}_1, \vec{k}_2; z\big) &= \Big(b^{\textrm{bod}}_4(z) - b^{\textrm{bod}}_1(z) + b^{\textrm{bod}}_2(z)\Big) + b^{\textrm{bod}}_1(z) \beta\big(\vec{k}_1, \vec{k}_2\big) \\
            &+ \bigg[b^{\textrm{bod}}_1(z) - \bigg(1 - \frac{a^{(2)}_{\gamma}(z)}{2}\bigg) b^{\textrm{bod}}_2(z)\bigg] \gamma\big(\vec{k}_1, \vec{k}_2\big)
        \end{aligned}
    \end{equation}
    \begin{equation}
        \begin{aligned}
            K^{(s)}_3\big(\vec{k}, \vec{q}, -\vec{q}; &z\big) = \frac{b^{\textrm{bod}}_1(z)}{3} \bigg[\beta\big(\vec{k} - \vec{q}, \vec{q}\big) \beta\big(\vec{k}, -\vec{q}\big) + \beta\big(\vec{k} + \vec{q}, -\vec{q}\big) \beta\big(\vec{k}, \vec{q}\big)\bigg] \\
            &
            \begin{aligned}
                + \frac{1}{6} \Big(d^{(2)}_{\gamma}(z) b^{\textrm{bod}}_1(z) &- \tilde{b}^{\textrm{bod}}_3(z) \big(d^{(2)}_{\gamma}(z) - a^{(3)}_{\gamma a}(z)\big) - 2 a^{(2)}_{\gamma}(z) \tilde{b}^{\textrm{bod}}_5(z)\Big) \\
                &\times \bigg[\gamma\big(\vec{k} - \vec{q}, \vec{q}\big) \gamma\big(\vec{k}, -\vec{q}\big) + \gamma\big(\vec{k} + \vec{q}, -\vec{q}\big) \gamma\big(\vec{k}, \vec{q}\big)\bigg]
            \end{aligned} \\
            &
            \begin{aligned}
                + \frac{1}{6} \Big(d^{(2)}_{\gamma}(z) b^{\textrm{bod}}_1(z) &+ \tilde{b}^{\textrm{bod}}_3(z) \big(2 h(z) - d^{(2)}_{\gamma}(z)\big)\Big) \\
                & \times \bigg[\beta\big(\vec{k} - \vec{q}, \vec{q}\big) \gamma\big(\vec{k}, -\vec{q}\big) + \beta\big(\vec{k} + \vec{q}, -\vec{q}\big) \gamma\big(\vec{k}, \vec{q}\big)\bigg].
            \end{aligned}
        \end{aligned}
        \label{eq:AppModelBiasedKernelThirdOrderDescendant}
    \end{equation}
\end{subequations}

Note that, usually, the fifth bias parameter, $\tilde{b}^{\textrm{bod}}_5$, does not appear in the literature, as it is degenerate with $\tilde{b}^{\textrm{bod}}_3$ in an EdS cosmology. This can be seen from eq.~\eqref{eq:AppModelBootstrapHFunction}, as $d^{(2)}_{\gamma, \textrm{EdS}} = 2 h_{\textrm{EdS}}$ leads to the removal of the $\tilde{b}^{\textrm{bod}}_3$ dependency in the second-to-last line. For cosmologies beyond EdS, the basis of descendants thus introduces five independent bias parameters, one more than the Galileons. 

In anticipation that the EdS approximation of the kernels is generally a good approximation~\cite{Bernardeau:2001qr}, a strong correlation between $\tilde{b}^{\textrm{bod}}_3$ and $\tilde{b}^{\textrm{bod}}_5$ should be expected. We therefore fix one of the parameters, thus enabling a direct comparison with the Galileons. Indeed, it is straightforward to show that by setting $\tilde{b}^{\textrm{bod}}_3 \equiv b^{\textrm{bod}}_1$ and introducing a new bias parameter $b^{\textrm{bod}}_3$ as
\begin{equation}
    b^{\textrm{bod}}_3(z) = \tilde{b}^{\textrm{bod}}_3(z) + \frac{2 a^{(2)}_{\gamma}(z)}{d^{(2)}_{\gamma}(z) - a^{(3)}_{\gamma a}(z)} \tilde{b}^{\textrm{bod}}_5(z),
\end{equation}
the BoD bias parameters are given in terms of the Galileons in the following way
\begin{subequations}
    \begin{equation}
        b^{\textrm{bod}}_1(z) = b_1(z)
    \end{equation}
    \begin{equation}
        b^{\textrm{bod}}_2(z) = b_1(z) + \frac{1}{1 - a^{(2)}_{\gamma}(z) / 2} b_{\mathcal{G}_2}(z)
    \end{equation}
    \begin{equation}
        b^{\textrm{bod}}_3(z) = b_1(z) + \frac{2}{d^{(2)}_{\gamma}(z) - a^{(3)}_{\gamma a}(z)} b_{\times}(z)
    \end{equation}
    \begin{equation}
        b^{\textrm{bod}}_4(z) = \frac{b_2(z)}{2} - \frac{1}{1 - a^{(2)}_{\gamma}(z)/2} b_{\mathcal{G}_2}(z).
    \end{equation}
\end{subequations}

In the EdS limit, these results agree with those of~\cite{Piga:2022mge} and also those that are implemented in \texttt{PyBird}.

\subsection{Biased Kernels in Redshift Space}\label{app:RSDKernels}

In redshift space, the kernels for biased tracers are given by (see, e.g.,~\cite{Amendola:2023awr})
\begin{subequations}
    \begin{equation}
        Z^{(s)}_1\big(\vec{k}_1; z\big) = K^{(s)}_1\big(\vec{k}_1; z\big) + f(z) \mu_1^2
    \end{equation}
    \begin{equation}
        \begin{aligned}
            Z^{(s)}_2\big(\vec{k}_1, \vec{k}_2; z\big) &= K^{(s)}_2\big(\vec{k}_1, \vec{k}_2; z\big) + f(z) \mu_{12}^2 G^{(s)}_2\big(\vec{k}_1, \vec{k}_2; z\big) \\
            &+ \frac{f(z) k_{12} \mu_{12}}{2} \bigg[\frac{\mu_2}{k_2} Z^{(s)}_1\big(\vec{k}_1; z\big) + \frac{\mu_1}{k_1} Z^{(s)}_1\big(\vec{k}_2; z\big)\bigg]
        \end{aligned}
        \label{eq:AppRSDKernelsZ2}
    \end{equation}
    \begin{equation}
        \begin{aligned}
            Z^{(s)}_3\big(\vec{k}_1, \vec{k}_2, \vec{k}_3; z\big) &= K^{(s)}_3\big(\vec{k}_1, \vec{k}_2, \vec{k}_3; z\big) + f(z) \mu_{123}^2 G^{(s)}_3\big(\vec{k}_1, \vec{k}_2, \vec{k}_3; z\big) \\
            &
            \begin{aligned}
            + \frac{f(z) k_{123} \mu_{123}}{3} &\bigg[\frac{\mu_3}{k_3} \Big(K^{(s)}_2\big(\vec{k}_1, \vec{k}_2; z\big) + f(z) \mu_{12}^2 G^{(s)}_2\big(\vec{k}_1, \vec{k}_2; z\big)\Big) \\
            &+ \frac{\mu_{12}}{k_{12}} G^{(s)}_2\big(\vec{k}_1, \vec{k}_2; z\big) Z^{(s)}_1\big(\vec{k}_3; z\big) + \cyc\bigg]
            \end{aligned}
            \\
            &+ \frac{f^2(z) k^2_{123} \mu^2_{123}}{6} \bigg[\frac{\mu_1}{k_1} \frac{\mu_2}{k_2} Z^{(s)}_1\big(\vec{k}_3; z\big) + \cyc\bigg].
        \end{aligned}
    \end{equation}
\end{subequations}

Similarly to eq.~\eqref{eq:AppModelBiasedKernelThirdOrder}, when considering $Z^{(s)}_3(\vec{k}, \vec{q}, -\vec{q})$, the remaining $b_2$ dependence in the second order kernel $K^{(s)}_2$ can be used to renormalise $f(z)$ at tree-level, leaving the remainder
\begin{equation}
    \begin{aligned}
        Z^{(s)}_3\big(\vec{k}, \vec{q}, -\vec{q}; z\big) &= K^{(s)}_3\big(\vec{k}, \vec{q}, -\vec{q}; z\big) + f(z) \mu^2 G^{(s)}_3\big(\vec{k}, \vec{q}, -\vec{q}; z\big) \\
        &
        \begin{aligned}
        + \frac{f(z) k \mu}{3} &\bigg[\frac{\mu_q}{q} \Big(K^{(s)}_2\big(\vec{k}, -\vec{q}; z\big)\Big\rvert_{b_2 = 0} + f(z) \mu_{\vec{k} - \vec{q}}^2 G^{(s)}_2\big(\vec{k}, -\vec{q}; z\big)\Big) \\
        &+ \frac{\mu_{\vec{k} - \vec{q}}}{|\vec{k} - \vec{q}|} G^{(s)}_2\big(\vec{k}, -\vec{q}; z\big) Z^{(s)}_1\big(\vec{q}; z\big) + (\vec{q} \leftrightarrow -\vec{q})\bigg]
        \end{aligned}
        \\
        &- \frac{f^2(z) k^2 \mu^2}{6} \frac{\mu_q^2}{q^2} Z^{(s)}_1\big(\vec{k}; z\big).
    \end{aligned}
\end{equation}

\section{IR Resummation in depth}\label{app:ir}

\subsection{Direct Method}\label{app:ir-direct}

The coefficients $\mathcal{Q}^{\ell \ell'}_{n n_0 \alpha || \mathcal{N}-j}$ in eq.~\eqref{eq:IRResumCoefQDirect} for the direct method of the IR resummation can be derived by closely following the approach of~\cite{Lewandowski:2015ziq}. In slight deviation, each exponential term in eq.~\eqref{eq:IRResumKernelK} for $K_0(\vec{k}, \vec{q})$ is expanded perturbatively straight away,
\begin{equation}
    \begin{aligned}
        K_0\big(\vec{k}, \vec{q}; z\big) &\approx \sum_{n_{0a} = 0}^{N_{0a}} \frac{(-1)^{n_{0a}}}{n_{0a}!} \bigg(\frac{1}{2} \Xi_0(q; z) k^2 \bigg)^{n_{0a}} \\
        &\times \sum_{n_{0b} = 0}^{N_{0b}} \frac{(-1)^{n_{0b}}}{n_{0b}!} \bigg(\frac{1}{2} \Xi_0(q; z) k^2 \mu^2 f(z)\big(2+f(z)\big)\bigg)^{n_{0b}} \\
        &\times \sum_{n_2 = 0}^{N_2} \frac{(-1)^{n_2}}{n_2!} \bigg(\frac{1}{2} \Xi_2(q; z) k^2 \big(\hat{k} \cdot \hat{q} + f(z)\mu\mu_q\big)^2\bigg)^{n_2},
    \end{aligned}
    \label{eq:AppIRResumKernelKExpandDirect}        
\end{equation}
while $K_0^{-1}\rvert\rvert_{\mathcal{N} - j}$ is simply given by
\begin{equation}
    \begin{aligned}
        K_0^{-1}\big(\vec{k}, \vec{q}; &z\big)\Big\rvert\Big\rvert_{\mathcal{N} - j} = \sum_{\kappa = 0}^{\mathcal{N}-j} \sum_{\substack{\kappa_{0a}, \kappa_{0b}, \kappa_2 \\ \sum_i \kappa_i = \kappa}} \frac{1}{\kappa_{0a}! \kappa_{0b}! \kappa_2!} \bigg(\frac{1}{2} \Xi_0(q; z) k^2 \bigg)^{\kappa_{0a}} \\
        &\times \bigg(\frac{1}{2} \Xi_0(q; z) k^2 \mu^2 f(z)\big(2+f(z)\big)\bigg)^{\kappa_{0b}} \bigg(\frac{1}{2} \Xi_2(q; z) k^2 \big(\hat{k} \cdot \hat{q} + f(z)\mu\mu_q\big)^2\bigg)^{\kappa_2}.
    \end{aligned}
\end{equation}

Remembering that eq.~\eqref{eq:AppIRResumKernelKExpandDirect} keeps the exponential terms strictly to orders $N_{0a}$, $N_{0b}$, and $N_2$, the two expressions are combined into $F_{||\mathcal{N}-j}(\vec{k}, \vec{q})$~\eqref{eq:IRResumKernelF} as
\begin{equation}
    \begin{aligned}
        F_{||\mathcal{N}-j}\big(\vec{k}, \vec{q}; z\big) &\approx \sum_{\kappa = 0}^{\mathcal{N}-j} \sum_{\substack{\kappa_{0a}, \kappa_{0b}, \kappa_2 \\ \sum_i \kappa_i = \kappa}} \frac{1}{\kappa_{0a}! \kappa_{0b}! \kappa_2!} \\
        &\times \sum_{n_{0a} = 0}^{N_{0a} - \kappa_{0a}} \frac{(-1)^{n_{0a}}}{n_{0a}!} \bigg(\frac{1}{2} \Xi_0(q; z) k^2 \bigg)^{n_{0a} + \kappa_{0a}} \\
        &\times \sum_{n_{0b} = 0}^{N_{0b} - \kappa_{0b}} \frac{(-1)^{n_{0b}}}{n_{0b}!} \bigg(\frac{1}{2} \Xi_0(q; z) k^2 \mu^2 f(z)\big(2+f(z)\big)\bigg)^{n_{0b} + \kappa_{0b}} \\
        &\times \sum_{n_2 = 0}^{N_2 - \kappa_2} \frac{(-1)^{n_2}}{n_2!} \bigg(\frac{1}{2} \Xi_2(q; z) k^2 \big(\hat{k} \cdot \hat{q} + f(z)\mu\mu_q\big)^2\bigg)^{n_2 + \kappa_2}.
    \end{aligned}
    \label{eq:AppIRResumKernelFExpandDirect}
\end{equation}

Consider now the $\hat{q}$ integral in eq.~\eqref{eq:IRResumKernelQ} and the last line of the previous expression,
\begin{align}
    \int d^2\hat{q} \, e^{-i \vec{k} \cdot \vec{q}} F_{||\mathcal{N}-j}\big(\vec{k}, \vec{q}; z\big) \mathcal{L}_{\ell'}(\mu_q) &\sim \int d^2\hat{q} \, e^{-i \vec{k} \cdot \vec{q}} \, (\hat{k} \cdot \hat{q} + f(z)\mu\mu_q)^{2 m} \mathcal{L}_{\ell'}(\mu_q) \nonumber \\
    &\;
    \begin{aligned}
        = \sum_{i = 0}^{2 m} &\binom{2 m}{i} f^i(z) \mu^i \\
        &\times \int d^2\hat{q} \, e^{-i \vec{k} \cdot \vec{q}} \big(\hat{k} \cdot \hat{q}\big)^{2 m - i} \mu_q^i \, \mathcal{L}_{\ell'}(\mu_q),
    \end{aligned}
\end{align}
where a binomial expansion was performed to get from the first line to the latter two lines. By following~\cite{Lewandowski:2015ziq}, the integral in the last line is given by
\begin{equation}
    \begin{aligned}
        \int d^2\hat{q} e^{-i \vec{k} \cdot \vec{q}} \big(\hat{k} \cdot \hat{q}\big)^{2 m - i} \mu_q^i \mathcal{L}_{\ell'}(\mu_q) = 4\pi \, (-i)^{\ell'} (-1)^{m} \sum_{j = -i}^{i} &(-i)^{j + i} \frac{2 (\ell' + j) + 1}{2} \\
        &\times x^{i}_{\ell', \ell' + j} \, j^{(2 m - i)}_{\ell' + j}(kq) \mathcal{L}_{\ell' + j}(\mu),
    \end{aligned}
\end{equation}
where $j^{(n)}_{\ell}(x)$ is the $n$'th derivative of the (spherical) Bessel function of order $\ell$, and
\begin{equation}
    x^{n}_{\ell \ell'} = \int_{-1}^1 d\mu \; \mu^n \mathcal{L}_{\ell}(\mu) \mathcal{L}_{\ell'}(\mu).
    \label{eq:AppIRResumDoubleLegendreIntegral}
\end{equation}

Inserting this result, together with eq.~\eqref{eq:AppIRResumKernelFExpandDirect}, into eq.~\eqref{eq:IRResumKernelQ} for $Q^{\ell \ell'}(k, q; z)_{||\mathcal{N}-j}(k,q)$, and using eq.~\eqref{eq:AppIRResumDoubleLegendreIntegral} for the remaining $\mu$ integral, we find
\begin{equation}
    \begin{aligned}
        Q^{\ell \ell'}_{||\mathcal{N}-j}(k, q; z) \approx \sum_{\kappa = 0}^{\mathcal{N}-j} (-1)^{\kappa} & \sideset{}{'}\sum_{\substack{\kappa_{0a}, \kappa_{0b}, \kappa_2 \\ \sum_i \kappa_i = \kappa}} \; \sideset{}{'}\sum_{n_{0a} = \kappa_{0a}}^{N_{0a}} (-1)^{n_{0a}} \sideset{}{'}\sum_{n_{0b} = \kappa_{0b}}^{N_{0b}} (-1)^{n_{0b}} \sideset{}{'}\sum_{n_2 = \kappa_2}^{N_2} (-1)^{n_2} \\
        &\times \bigg(\frac{1}{2} \Xi_0(q; z) k^2 \bigg)^{n_{0a} + n_{0b}} \bigg(\frac{1}{2} \Xi_2(q; z) k^2\bigg)^{n_2} J^{\ell \ell'}_{n_{0b} n_2}(kq; z),
    \end{aligned}
    \label{eq:AppIRResumQCoefficientsPre}
\end{equation}
where index shifts $n_i \to n_i - \kappa_i$ have been performed relative to eq.~\eqref{eq:AppIRResumKernelFExpandDirect}, where the notation $(\sum')_{i = i_0} = \sum_i \frac{1}{(i - i_0)!}$ has been used, and where the coefficients $J^{\ell \ell'}_{n m}(kq; z)$ have been defined as
\begin{equation}
    \begin{aligned}
        J^{\ell \ell'}_{nm}(kq; z) &= \frac{2 \ell + 1}{2} (-1)^m \Big[f(z) \big(2+f(z)\big)\Big]^n \\
        &\times\sum_{i = 0}^{2 m} \binom{2m}{i} f^i(z) \sum_{j = -i}^i (-i)^{j + i} \, \frac{2 (\ell' + j) + 1}{2} \, x^{i}_{\ell', \ell' + j} \, x^{2n + i}_{\ell, \ell' + j} \, j^{(2m - i)}_{\ell' + j}(kq).
    \end{aligned}
    \label{eq:AppIRResumJCoefficientsPre}
\end{equation}

Lastly, by using that the first derivative of the $\ell$'th order (spherical) Bessel function can be expressed as (see eq. 10.1.20 in~\cite{AbramowitzStegun})
\begin{equation}
    \frac{d}{dx} j_{\ell}(x) = \frac{\ell}{2\ell + 1} \, j_{\ell - 1}(x) - \frac{\ell + 1}{2\ell + 1} j_{\ell + 1}(x),
    \label{eq:AppIRResumBesselFirstDerivative}
\end{equation}
the $n$'th derivative of the $\ell$'th order Bessel fucntion, $j^{(n)}_{\ell}$, can hence be expressed as a sum of Bessel functions of various orders,
\begin{equation}
    j^{(n)}_{\ell}(x) = \sum_{\ell' = \max(0, \ell - n)}^{\ell + n} a^{n}_{\ell \ell'} j_{\ell'}(x) \, ,
    \label{eq:AppIRResumBesselDerivative}
\end{equation}
where the coefficients $a^{n}_{\ell \ell'}$ are non-trivial for even $\ell + \ell' + n$ and $\max(0, \ell - n) \leq \ell' \leq \ell + n$, and can be determined recursively from the initial condition $a^0_{\ell \ell} = 1$ via
\begin{equation}
    a^n_{\ell \ell'} = \frac{\ell' + 1}{2 \ell' + 3} a^{n-1}_{\ell \ell'+1} - \frac{\ell'}{2 \ell' - 1} a^{n-1}_{\ell \ell'-1} \, .
    \label{eq:AppIRResumBesselDerivativeCoef}
\end{equation}
We can then write $J^{\ell \ell'}_{n m}(kq)$ also as
\begin{equation}
    J^{\ell \ell'}_{n m}(kq; z) = \sum_{\alpha = A_i(\ell', m)}^{A_f(\ell', m)} \mathcal{J}^{\ell \ell'}_{n m \alpha}(z) j_\alpha(kq) \, ,
    \label{eq:AppIRResumJCoefficients}
\end{equation}
where $A_i(\ell', m) = \max\{0, \ell' - 2m\}$ and $A_f(\ell', m) = \ell' + 2m$ follow from the bounds of the sum in eq.~\eqref{eq:AppIRResumBesselDerivative}, and with $\mathcal{J}^{\ell \ell'}_{n m \alpha}(z)$ as defined in eq.~\eqref{eq:IRResumCoefJ}. Inserting this expression into eq.~\eqref{eq:AppIRResumQCoefficientsPre} and extracting powers of $\Xi_0(q) k^2$ and $\Xi_2(q) k^2$, we arrive at
\begin{equation}
    \begin{aligned}
        Q^{\ell \ell'}_{||\mathcal{N}-j}(k, q; z) = \sum_{n = 0}^{N} k^{2n} \sum_{n_0 = N_{0i}(n)}^{N_{0f}(n)} \sum_{\alpha = A_i(\ell', m)}^{A_f(\ell', m)} &\mathcal{Q}^{\ell \ell'}_{n n_0 \alpha || \mathcal{N}-j}(z) \\
        &\times \Xi_0^{n_0}(q; z) \Xi_2^{n-n_0}(q; z) j_{\alpha}(kq) \,,
    \end{aligned}
    \label{eq:AppIRResumQCoefficients}
\end{equation}
with $\mathcal{Q}^{\ell \ell'}_{n n_0 \alpha || \mathcal{N}-j}(z)$ as defined in eq.~\eqref{eq:IRResumCoefQDirect}, and where $N_{0i}(n) = \max(0, n - N_2)$ and $N_{0f}(n) = \min(n, N_{0a} + N_{0b})$\footnote
{
Letting $n_0 = n_{0a} + n_{0b}$ be variable for fixed $n = n_0 + n_{2}$ in eq.~\eqref{eq:AppIRResumQCoefficientsPre}, the allowed range for $n_0$ can be derived as follows. Apart from the condition that $n_0 \geq 0$, another condition follows from $n_0 = n - n_2 \geq n - N_2$. The smallest value that $n_0$ can attain is thus $N_{0i}(n) \equiv \max(0, n - N_2)$. Similarly, while $n_0 = n - n_2 \leq n$, also $n_0 = n_{0a} + n_{0b} \leq N_{0a} + N_{0b}$ must hold. Therefore, the largest value of $n_0$ is given by $N_{0f}(n) \equiv \min(n, N_{0a} + N_{0b})$.
}, such that, together with eq.~\eqref{eq:IRResumPowerSpectrum} for the resummed power spectrum, $P^{\ell}(k)\rvert_{\mathcal{N}}$, the expression in eq.~\eqref{eq:IRResumPowerSpectrumDirect} is recovered.\footnote
{
Note that the term in eq.~\eqref{eq:AppIRResumQCoefficients} with $n = 0$ forces $n_0 = 0$, since $N_{0i}(0) = 0 = N_{0f}(0)$, such that
\[
    \mathcal{Q}^{\ell \ell'}_{0 0 \alpha || \mathcal{N}-j}(z) = \mathcal{J}^{\ell \ell'}_{0 0 \alpha}(z) = \frac{2\ell + 1}{2} \frac{2\ell' + 1}{2} \underbrace{x^{0}_{\ell' \ell'}}_{\frac{2}{2\ell' + 1}} \underbrace{x^{0}_{\ell \ell'}}_{\frac{2}{2\ell' + 1} \delta_{\ell \ell'}} \underbrace{a^{0}_{\ell' \alpha}}_{\delta_{\ell' \alpha}} = \delta_{\ell \ell'} \delta_{\ell' \alpha}
\]
leaving $Q^{\ell \ell'}_{||\mathcal{N}-j}(k, q; z) \sim \delta_{\ell \ell'} j_{\ell}(kq)$, and thus
\[
    P^{\ell}(k; z) \Big\rvert_{\mathcal{N}} \sim 4\pi (-i)^{\ell} \sum_{j = 0}^{\mathcal{N}} \int_0^{\infty} dq \; q^2 \xi^{\ell}(q; z)_j \, j_{\ell}(kq) = \sum_{j = 0}^{\mathcal{N}} P^{\ell}(k; z)_j = P^{\ell}(k; z)\Big\rvert\Big\rvert_{\mathcal{N}} \, .
\]
Importantly, this result is also the asymptotic behaviour of the IR resummation for $k \to 0$, i.e., the IR resummed power spectrum approaches the unresummed power spectrum.
}

\subsection{Asymptotic Method} \label{app:ir-asympt}

The coefficients $\tilde{\mathcal{Q}}^{\ell \ell'}_{n \tilde{n} \tilde{n}_0 \alpha || \mathcal{N}-j}$~\eqref{eq:IRResumCoefQAsymptote} for the asymptotic method of the IR resummation follow similarly. The main difference comes from the different treatment of $K_0(\vec{k}, \vec{q})$. Starting from eq.~\eqref{eq:IRResumKernelKAsymptote}, we expand only the last two exponentials,
\begin{align}
    K_0\big(\vec{k}, \vec{q}; z\big) &\approx \exp\bigg[-k^2 q^2 \Xi^{(2)}(z)\bigg] \label{eq:AppIRResumKernelKExpandAsymptote} \\
    &\times \sum_{n_{0} = 0}^{N_{0}} \frac{(-1)^{n_{0}}}{n_{0}!} \bigg(\frac{1}{2} k^2 q^2 \Big[\tilde{\Xi}_0(q; z) \Big(1 + f(z) \big(2 + f(z)\big)\Big) - \Xi^{(2)}_0(z) \big(1 + f(z)\big)^2 \Big] \bigg)^{n_{0}} \nonumber \\
    &\times \sum_{n_2 = 0}^{N_2} \frac{(-1)^{n_2}}{n_2!} \bigg(\frac{1}{2} k^2 q^2 \Big[\tilde{\Xi}_2(q; z) \big(\hat{k} \cdot \hat{q} + f(z)\mu\mu_q\big)^2 - \Xi^{(2)}_2(z) \big(1 + f(z)\big)^2\Big]\bigg)^{n_2}. \nonumber
\end{align}

By further performing Binomial expansions and combining with \begin{equation}
    \begin{aligned}
        K_0^{-1}\big(\vec{k}, \vec{q}; &z\big)\Big\rvert\Big\rvert_{\mathcal{N} - j} = \sum_{\kappa = 0}^{\mathcal{N}-j} \bigg(\frac{1}{2} k^2 q^2\bigg)^{\kappa} \sum_{\substack{\kappa_0, \kappa_2 \\ \sum_i \kappa_i = \kappa}} \frac{1}{\kappa_0! \kappa_2!} \\
        &\times \bigg(\tilde{\Xi}_0(q; z) \Big(1 + \mu^2 f(z)\big(2+f(z)\big)\Big)\bigg)^{\kappa_0} \bigg(\tilde{\Xi}_2(q; z) \big(\hat{k} \cdot \hat{q} + f(z)\mu\mu_q\big)^2\bigg)^{\kappa_2} \, ,
    \end{aligned}
\end{equation}
we find for $F_{||\mathcal{N}-j}(\vec{k}, \vec{q})$,
\begin{equation}
    \begin{aligned}
        F_{||\mathcal{N}-j}\big(\vec{k}, \vec{q}; z\big) \, & \approx \exp\bigg[-k^2 q^2 \Xi^{(2)}(z)\bigg] \\
        &\times \sum_{\kappa = 0}^{\mathcal{N}-j} (-1)^{\kappa} \sideset{}{'}\sum_{\substack{\kappa_{0}, \kappa_2 \\ \sum_i \kappa_i = \kappa}} \left(\prod_{i \in \{0, 2\}} \, \sideset{}{'}\sum_{n_i = \kappa_i}^{N_i} \; \sideset{}{''}\sum_{m_i = \kappa_i}^{n_i} \right) \bigg(\frac{1}{2} k^2 q^2\bigg)^{n_0 + n_2} (-1)^{m_0 + m_2} \\
        &\times \bigg(\tilde{\Xi}_0(q; z) \Big(1 +  \mu^2 f(z)\big(2+f(z)\big)\Big)\bigg)^{m_0} \bigg(\Xi^{(2)}_0(z) \big(1 + f(z)\big)^2\bigg)^{n_{0} - m_0} \\
        &\times \bigg(\tilde{\Xi}_2(q; z) \big(\hat{k} \cdot \hat{q} + f(z)\mu\mu_q\big)^2\bigg)^{m_2} \bigg(\Xi^{(2)}_2(z) \big(1 + f(z)\big)^2\bigg)^{n_2 - m_2} \, ,
    \end{aligned}
    \label{eq:AppIRResumKernelFExpandAsymptote}
\end{equation}
where we used the notation $(\sum'')_{i = i_0}^{i_1} = \sum_{i = i_0}^{i_1} \binom{i_1 - i_0}{i - i_0}$. The angular integrals of eq.~\eqref{eq:IRResumKernelQ} can now be done similarly as for the direct method, ultimately leading to
\begin{equation}
    \begin{aligned}
        Q^{\ell \ell'}_{||\mathcal{N}-j}(k, q; z) \, & \approx \exp\bigg[-k^2 q^2 \Xi^{(2)}(z)\bigg] \\
        &\times \sum_{\kappa = 0}^{\mathcal{N}-j} (-1)^{\kappa} \sideset{}{'}\sum_{\substack{\kappa_{0}, \kappa_2 \\ \sum_i \kappa_i = \kappa}} \left(\prod_{i \in \{0, 2\}} \, \sideset{}{'}\sum_{n_i = \kappa_i}^{N_i} \; \sideset{}{''}\sum_{m_i = \kappa_i}^{n_i} \right) \bigg(\frac{1}{2} k^2 q^2\bigg)^{n_0 + n_2} (-1)^{m_0 + m_2} \\
        &\times \bigg(\Xi^{(2)}_0(z) \big(1 + f(z)\big)^2\bigg)^{n_{0} - m_0}  \bigg(\Xi^{(2)}_2(z) \big(1 + f(z)\big)^2\bigg)^{n_2 - m_2} \\
        &\times \tilde{\Xi}^{m_0}_0(q; z) \tilde{\Xi}^{m_2}_2(q; z) \tilde{J}^{\ell \ell'}_{m_0 m_2}(kq) \, ,
    \end{aligned}
    \label{eq:AppIRResumQCoefficientsAsymptotePre}
\end{equation}
where\footnote{Notice that, due to the $+1$ in the $1 + \mu^2 f(2+f)$ term of eq.~\eqref{eq:AppIRResumKernelFExpandAsymptote}, the coefficients $\tilde{J}^{\ell \ell'}_{n m}$ are related to the coefficients $J^{\ell \ell'}_{n m}$ via $\tilde{J}^{\ell \ell'}_{n m} = \sum_{\tilde{n} = 0}^n \binom{n}{\tilde{n}} J^{\ell \ell'}_{\tilde{n} m}$, as can be seen by comparing eq.~\eqref{eq:AppIRResumJCoefficientsPre} with eq.~\eqref{eq:AppIRResumJCoefficientsAsymptotePre}.}
\begin{equation}
    \begin{aligned}
        \tilde{J}^{\ell \ell'}_{m_0 m_2}(kq; z) &= \frac{2 \ell + 1}{2} (-1)^{m_2} \sideset{}{''}\sum_{i_0 = 0}^{m_0} \Big[f(z) \big(2+f(z)\big)\Big]^{i_0} \sideset{}{''}\sum_{i_2 = 0}^{2 m_2} f^{i_2}(z) \\
        &\times \sum_{j_2 = -i_2}^i (-i)^{j_2 + i_2} \, \frac{2 (\ell' + j_2) + 1}{2} \, x^{i}_{\ell', \ell' + j_2} \, x^{2i_0 + i_2}_{\ell, \ell' + j_2} \, j^{(2m_2 - i_2)}_{\ell' + j_2}(kq).
    \end{aligned}
    \label{eq:AppIRResumJCoefficientsAsymptotePre}
\end{equation}

Finally, extracting Bessel functions according to eq.~\eqref{eq:AppIRResumBesselDerivative}, powers of $(kq)^2$, and powers of $\tilde{\Xi}_0(q)$ and $\tilde{\Xi}_2(q)$, the coefficients for the asymptotic method of the IR resummation become,
\begin{equation}
    \begin{aligned}
        Q^{\ell \ell'}_{||\mathcal{N}-j}(k, q; z) & \approx \sum_{n = 0}^{N} (kq)^{2n} \sum_{\tilde{n} = 0}^{n} \sum_{\tilde{n}_0 = N_{0i}(\tilde{n})}^{N_{0f}(\tilde{n})} \sum_{\alpha = A_i(\ell', \tilde{n}_0)}^{A_f(\ell', \tilde{n}_0)} \tilde{\mathcal{Q}}^{\ell \ell'}_{n \tilde{n} \tilde{n}_0 \alpha || \mathcal{N}-j}(z) \\
        &\times \exp\bigg[-k^2 q^2 \Xi^{(2)}(z)\bigg] \tilde{\Xi}_0^{\tilde{n}_0}(q; z) \tilde{\Xi}_2^{\tilde{n}-\tilde{n}_0}(q; z) j_{\alpha}(kq) \, ,
    \end{aligned}
    \label{eq:AppIRResumQCoefficientsAsympt}
\end{equation}
where the coefficients $\tilde{\mathcal{Q}}^{\ell \ell'}_{n \tilde{n} \tilde{n}_0 \alpha || \mathcal{N}-j}(z)$ are as defined in eq.~\eqref{eq:IRResumCoefQAsymptote}.

\section{Hypergeometric Functions}\label{app:Hyper}

\subsection{The Confluent Hypergeometric Function}\label{app:HyperConfluent}

The confluent hypergeometric function, ${}_1F_1(a; b; x)$, often also encountered as Kummer's function of the first kind, $M(a, b, x)$, is given by (see eq. 13.1.2 in~\cite{AbramowitzStegun})
\begin{equation}
    {}_1F_1(a; b; x) = \sum_{n = 0}^{\infty} \frac{(a)_n}{(b)_n} \frac{x^n}{n!} \, ,
    \label{eq:AppHypConfluent}
\end{equation}
where we consider $a \in \mathbb{C}$, $b \in \mathbb{R}^{+}$
, and $x \in \mathbb{R}$, and where $(a)_n$ is the rising Pochhammer symbol,
\begin{equation}
    (a)_n = \left\{
    \begin{array}{cc}
        1 & n = 0 \\
        a (a + 1) \cdots (a + n - 1) & n > 0
    \end{array}\right..
\end{equation}

Despite the series having an infinite radius of convergence, a numerical evaluation becomes challenging whenever $|a|$ or $|x|$ is large compared to $b$. As more terms are needed for the series to be safely truncated, not only does convergence require more computation time, but numerical errors also start to accumulate, rendering the final result potentially unusable. One is thus forced to find workarounds for an accurate computation of ${}_1F_1(a; b; x)$ for a large range of values for $a$, $b$, and $x$.

For values $|a| |x| < b$, the series~\eqref{eq:AppHypConfluent} converges sufficiently quickly. To determine convergence, we require the ratio of the $n$'th summand of the series, $T_n = (a)_n/(b)_n x^n/n!$, to the $n$'th partial sum, $S_n = \sum_{m = 0}^n T_m$, to be sufficiently small, $|T_n / S_n| < \epsilon = 10^{-10}$.

For values $|x| > 100$, we make use of the asymptotic expansions (see eq. 13.5.1 in~\cite{AbramowitzStegun})
\begin{subequations}
    \label{eqs:AppHypConfluentAsympt}
    \begin{align}
        {}_1F_1(a; b; x) &\underset{x \to \infty}{\sim} e^x x^{a-b} \frac{\Gamma(b)}{\Gamma(a)} \left[\sum_{n = 0}^{N} \frac{(b-a)_n (1-a)_n}{n!} x^{-n} + \mathcal{O}\big(x^{-(N+1)}\big) \right] \\
        {}_1F_1(a; b; x) &\underset{x \to -\infty}{\sim} (-x)^{-a} \frac{\Gamma(b)}{\Gamma(b-a)} \left[\sum_{n = 0}^{N} \frac{(a)_n (1+a-b)_n}{n!} (-x)^{-n} + \mathcal{O}\big(x^{-(N+1)}\big) \right].
    \end{align}
\end{subequations}

As the sums are non-convergent for $N \to \infty$, in addition to the convergence criterion of the series~\eqref{eq:AppHypConfluent}, we also truncate the asymptotic expansions prematurely if $|T_n| > |T_{n-1}|$, where $T_n$ is the $n$'th summand of the respective sum, i.e., when the sum starts to diverge. Furthermore, we require the remainders,
\begin{subequations}
    \begin{align}
        \mathcal{O}\big(x^{-N}\big) &\underset{x \to \infty}{=} \frac{(b-a)_N (1-a)_N}{N!} x^{-N} \left[ \frac{2}{3} - b + 2a + x - N + \mathcal{O}\big(x^{-1}\big) \right] \\
        \mathcal{O}\big(x^{-N}\big) &\underset{x \to -\infty}{=} \frac{(a)_N (1+a-b)_N}{N!} (-x)^{-N} \left[ \frac{3}{4} + \frac{1 + 2b -4a - 2N}{8x} + \mathcal{O}\big(x^{-2}\big) \right] \, ,
    \end{align}
\end{subequations}
to differ from the last summand, $T_{N-1}$, by at most $10\%$ to categorise the series as converged.

Another approach to dealing with too large values of $|x|$, but also too large values of $|a|$, is to use recurrence relations (see eq. 13.4.2 in~\cite{AbramowitzStegun}),
\begin{equation}
    {}_1F_1(a; b; x) = \bigg(1 + \frac{z}{b}\bigg) {}_1F_1(a; b + 1; x) - \frac{x}{b}\bigg(1 - \frac{a}{b+1}\bigg) {}_1F_1(a; b + 2; x).
    \label{eq:AppHypConfluentRecursion}
\end{equation}

The idea is to increase $b$ until $b \sim |a| |x|$ by recursively applying this relation, such that eq.~\eqref{eq:AppHypConfluent} can be used reliably for the resulting hypergeometric functions at the end of the recursion,
\begin{equation}
    {}_1F_1(a; b; x) = \alpha_n(a, b, x) {}_1F_1(a; b + 2^n; x) + \beta_n(a, b, x) {}_1F_1(a; b + 2^{n+1}; x),
    \label{eq:AppHypConfluentExpRecursion}
\end{equation}
where $\alpha_n$ and $\beta_n$ are defined recursively as
\begin{subequations}
    \begin{equation}
        \begin{aligned}
            \alpha_n(a, b, x) &= \alpha_{n-1}(a, b, x) \bigg[\alpha_{n-1}(a, b + 2^{n-1}, x) + \frac{\beta_{n-1}(a, b + 2^{n-1}, x)}{\alpha_{n-1}(a, b + 2^n, x)}\bigg] \\
            &+ \beta_{n-1}(a, b, x)
        \end{aligned}
    \end{equation}
    \begin{equation}
        \beta_n(a, b, x) = - \frac{\alpha_{n-1}(a, b, x) \beta_{n-1}(a, b + 2^{n-1}, x) \beta_{n-1}(a, b + 2^n, x)}{\alpha_{n-1}(a, b + 2^n, x)}
    \end{equation}
\end{subequations}
for $n > 1$ and with the initial conditions $\alpha_1(a, b, x) = 1 + \frac{x}{b}$ and $\beta_1(a, b, x) = - \frac{x}{b} (1 - \frac{a}{b + 1})$. We determine the depth of the recursion, $n$, from the condition $b + 2^n \geq |a| |x|$, i.e., 
\begin{equation}
    n = \bigg\lceil \log_2 \Big( - b + |a| |x|\Big) \bigg\rceil.
\end{equation}

This approach is used when neither the series~\eqref{eq:AppHypConfluent} nor the asymptotic expansions~\eqref{eqs:AppHypConfluentAsympt} are applicable, as well as when the asymptotes do not converge.

To validate our implementation of the confluent hypergeometric function, we first draw 100 random numbers for real $a \in (-50, 50)$, another 100 random numbers for $b \in (0, 50)$, 50 random numbers for positive values of $x$ via $\log_{10} x \in (-3, 2)$, and another 50 random numbers for negative values of $x$ via $\log_{10} (-x) \in (-3, 3)$, and compare the results of the $10^6$ different combinations with the implementation of \texttt{mpmath}~\cite{mpmath}. After repeating this process 10 times, we find an average relative error of $1.98 \times 10^{-12}$ and a maximal relative error of $1.01 \times 10^{-7}$.

Similarly, when performing the same test for complex values of $a$, i.e, drawing 100 random numbers for $a$ in the square $\Re(a) \in (-50, 50)$ and $\Im(a) \in (-70, 70)$, we find an average relative error of $1.24 \times 10^{-12}$ and a maximal relative error of $3.07 \times 10^{-10}$.

Note that, while extremely accurate for the parameter ranges tested above, the method of eq.~\eqref{eq:AppHypConfluentExpRecursion} is fairly slow when compared to the direct evaluation of eq.~\eqref{eq:AppHypConfluent}, especially for large recursion depths $n$.

\subsection{The Gaussian Hypergeometric Function}\label{app:HyperGauss}

Similarly to the confluent hypergeometric function, the Gaussian hypergeometric function can also be defined through a power series, also called the Gauss series (see eq.~15.1.1 in~\cite{AbramowitzStegun}),
\begin{equation}
    {}_2F_1(a, b; c; x) = \sum_{n = 0}^{\infty} \frac{(a)_n (b)_b}{(c)_n} \frac{x^n}{n!} \, ,
    \label{eq:AppHypGauss}
\end{equation}
where we consider $a,b \in \mathbb{C}$, $c \in \mathbb{C} \setminus \mathbb{Z}^{-}$, and $x \in \mathbb{R}$. Contrary to the confluent hypergeometric function, the Gauss series only has a radius of convergence of $|x| < 1$. However, since $x \equiv \eta = 1 - (hd)^2 \in (-\infty, 1)$\footnote{Note that $\eta = 1$ is excluded as either $h = 0$ or $d = 0$ is implied, both of which are unphysical, given that eq.~\eqref{eq:APEffectPowerSpectrumMultipole} breaks down.} is needed for the AP effect (cf. eq.~\eqref{eq:APEffectTransformationMatrix}), the Gauss series must be analytically continued to cover all possible values of $\eta$. This can be done by using the following transformation rules (see eqs.~15.3.3--15.3.9 in~\cite{AbramowitzStegun}),
\begin{subequations}
    \begin{align}
        &
        \begin{aligned}
            {}_2F_1(a, b; c; x) &= (1 - x)^{-a} \frac{\Gamma(c) \Gamma(b - a)}{\Gamma(b) \Gamma(c - a)} {}_2F_1\bigg(a, c - b; a - b + 1; \frac{1}{1 - x}\bigg) \\
            &+ (1 - x)^{-b} \frac{\Gamma(c) \Gamma(a - b)}{\Gamma(a) \Gamma(c - b)} {}_2F_1\bigg(b, c - a; b - a + 1; \frac{1}{1 - x}\bigg)\,,
        \end{aligned} \label{eq:AppHypGauss1} \\
        &{}_2F_1(a, b; c; x) = (1 - x)^{-a} {}_2F_1\bigg(a, c - b; c; \frac{x}{x - 1}\bigg)\,, \label{eq:AppHypGauss2} \\
        &
        \begin{aligned}
            {}_2F_1(a, b; c; x) &= \frac{\Gamma(c) \Gamma(c - a - b)}{\Gamma(c - a) \Gamma(c - b)} {}_2F_1\big(a, b; a + b - c + 1; 1 - x\big) \\
            &+ (1 - x)^{c - a - b} \frac{\Gamma(c) \Gamma(a + b - c)}{\Gamma(a) \Gamma(b)} {}_2F_1\big(c - a, c - b; c - a - b + 1; 1 - x\big)\,.
        \end{aligned} \label{eq:AppHypGauss3}
    \end{align}
\end{subequations}

By making use of these transformation rules as explained below, we only need to compute hypergeometric functions in the range $x \in (0, 1)$, such that Gauss' series~\eqref{eq:AppHypGauss} can be used. To determine convergence, we use the method of~\cite{Michel:2008}, and attempt to reach a relative error of $10^{-10}$.

For $x < 0$, we make use of eq.~\eqref{eq:AppHypGauss1} and eq.~\eqref{eq:AppHypGauss2}. Instead of testing which function to use for a given set of parameters, we first attempt to use eq.~\eqref{eq:AppHypGauss2}. If numerical errors larger than the tolerance are detected, eq.~\eqref{eq:AppHypGauss1} is probed as well. The final result is then chosen as the one with the smallest error.

We proceed similarly for $x \in [0, 1)$, for which we use eq.~\eqref{eq:AppHypGauss} and eq.~\eqref{eq:AppHypGauss3}. In this case, we start by probing eq.~\eqref{eq:AppHypGauss} for convergence according to the tolerance threshold and switch to eq.~\eqref{eq:AppHypGauss3} if this fails. The final result is then also chosen based on the smallest error.

One thing to note, though, is that eqs.~\eqref{eq:AppHypGauss1} and~\eqref{eq:AppHypGauss3} cannot be applied if $a - b \in \mathbb{Z}$ and $c - a - b \in \mathbb{Z}$, respectively. From eq.~\eqref{eq:APEffectTransformationMatrix}, this means that $\nu_m$, and hence the FFTLog bias $\nu$, can neither be integers nor half-integers, i.e., $\nu_m \neq k / 2$ for $k \in \mathbb{Z}$, which is generally never a problem.\footnote{For instance, this is precisely the reason the bias $\nu = 0.1$ was chosen for the FFTLog of the IR resummed part of the power spectrum $P^{\ell}_{\textrm{IR}}$ in table~\ref{tab:fftlog}, instead of the slightly more preferable one $\nu = 0$.} The only exceptions are the stochastic terms (cf. eq.~\eqref{eq:Stochasticity}), as they go with $\sim k^{2n}$. Hence, eq.~\eqref{eq:AppHypGauss3} cannot be used in the range $z \in (1/2, 1)$, for which the Gauss series is used instead.

To validate our implementation of the Gaussian hypergeometric function, we focus on the form appearing in eq.~\eqref{eq:AppInt-1} for the AP effect, ${}_2F_1(\nu, \mu + 1/2; \mu + 3/2; x)$. We first draw 500 random samples for $\nu \in (-50, 50)$, 100 random samples for $x \in (-10, 1)$, as well as use $\mu \in \{0, 1, \dots, 19\}$. We then compare the results of the $10^6$ different combinations with the implementation of ${}_2F_1$ in \texttt{mpmath}~\cite{mpmath}, and repeat this process 10 times, giving an average relative error of $2.08 \times 10^{-10}$ and a maximal relative error of $1.47 \times 10^{-8}$.

The same process is repeated for complex values of $\nu$, choosing 500 random numbers in the square $\Re(\nu) \in (-50, 50)$ and $\Im(\nu) \in (-100, 100)$, we then find an average relative error of $8.08 \times 10^{-11}$ and a maximal relative error of $5.94 \times 10^{-8}$.

Note that these ranges are well beyond what would be needed for the AP effect. For instance, $\Im(\nu) \sim \pm 100$ roughly correspond to the highest frequencies in the FFTLog of $P^{(0)}(k)$ of size $N \sim 512$ for the one-loop power spectrum or $N \sim 1024$ for the tree-level power spectrum, i.e., double/quadrupole the sizes we use by default (cf.~table~\ref{tab:fftlog}).

\section{Table of Integrals}

Here, we provide a collection of definite integrals that are used throughout the paper. Note that we often only cite the parameter spaces that are of interest for this work. See the cited references for the entire parameter spaces instead.

\begin{enumerate}
    \item For $\mu \in \mathbb{N}_0$, $\alpha \in \mathbb{R}$, and $\nu \in \mathbb{C}$ (see eq. 3.194(1) in~\cite{Gradshteyn_2007}),
\begin{equation}
    I_1(\alpha, \mu, \nu) \equiv \int_{-1}^{1} dx \frac{x^{2\mu}}{\big(1 - \alpha x^2\big)^\nu} = \frac{1}{\mu + 1/2} \, {}_2F_1\Big(\nu, \mu + 1/2; \mu + 3/2; \alpha\Big).
    \label{eq:AppInt-1}
\end{equation}

\item For $\alpha \in \mathbb{R}^{+}$ and $\nu \in \mathbb{C}$ with $\Re(\nu) > -1$ (see eq. 3.381(4) in~\cite{Gradshteyn_2007}),
\begin{equation}
    I_2(\alpha, \nu) \equiv \int_0^{\infty} dx \, x^{\nu} e^{-\alpha x^2} = \frac{1}{2} \alpha^{-(1+\nu)/2} \, \Gamma\left(\frac{\nu+1}{2}\right).
    \label{eq:AppIntExpPow}
\end{equation}
    
\item For $\mu \in \mathbb{N}_0$ and $\nu \in \mathbb{C}$ with $-\mu-1 < \Re(\nu) < 1$ (see eq. 6.561(14) in~\cite{Gradshteyn_2007}),
\begin{equation}
    I_3(\mu, \nu) \equiv \int_0^{\infty} dx \, x^{\nu} j_{\mu}(x) = \sqrt{\pi} \, 2^{\nu - 1} \frac{\Gamma(1/2 + \mu/2 + \nu/2)}{\Gamma(1 + \mu/2 - \nu/2)}.
    \label{eq:AppIntJnPow}
\end{equation}
    
\item For $\mu \in \mathbb{N}_0$, $\alpha \in \mathbb{R}^{+}$, and $\nu \in \mathbb{C}$ with $\Re(\nu) > -\mu-1$ (see eq. 6.631(1) in~\cite{Gradshteyn_2007}),
\begin{equation}
    \begin{aligned}
        I_4(\alpha, \mu, \nu) \equiv \int_0^{\infty} x^{\nu} e^{-\alpha x^2} j_{\mu}(x) &= \frac{\sqrt{\pi}}{2^{\mu + 2} a^{\frac{1}{2} (1 + \mu + \nu)}} \frac{\Gamma(1/2 + \mu/2 + \nu/2)}{\Gamma(\mu + 3/2)} \\
        &\times {}_1F_1\left(\frac{1 + \mu + \nu}{2}; \mu + \frac{3}{2}; - \frac{1}{4 \alpha}\right).
    \end{aligned}
    \label{eq:AppIntJnExpPow}
\end{equation}
\end{enumerate}

\section*{Acknowledgements}

LA acknowledges support by DFG under Germany’s Excellence Strategy EXC 2181/1 - 390900948 (the Heidelberg STRUCTURES Excellence Cluster), under Project 554679582 ``GeoGrav: Cosmological Geometry and Gravity with nonlinear physics'', and  by Tamkeen under the NYUAD Research Institute grant CG014. TSP is supported by Fundação Araucária (NAPI Fenômenos Extremos do Universo, grant 347/2024 PD\&I), FAPERJ project E-26/204.633/2024, and CAPES. MQ is supported by the Brazilian research agencies FAPERJ project E-26/201.237/2022 and CNPq (Conselho Nacional de Desenvolvimento Científico e Tecnológico). We acknowledge the use of the computational resources of the joint CHE / Milliways cluster, supported by a FAPERJ grants E-26/210.130/2023 and E-26/210.799/2025.

\bibliographystyle{JHEP} 
\bibliography{biblio}

\end{document}